\documentclass[twocolumn,showpacs,preprintnumbers,amsmath,amssymb]{revtex4}

\usepackage{graphicx,subfigure,color}
\usepackage{dcolumn}
\usepackage{bm}

\newcommand{\bra}[1]{\langle #1|}
\newcommand{\ket}[1]{|#1\rangle}

\def\tr{ \mbox{tr} }
\def\Tr{ \mbox{Tr} }

\begin{document}

\preprint{APS/123-QED}

\title{Quantum Mutual Information Along Unitary Orbits}

\author{Sania Jevtic, David Jennings and Terry Rudolph}
 \affiliation{Controlled Quantum Dynamics Theory, Department of Physics, Imperial College London, London SW7 2AZ}


\date{\today}

\begin{abstract}
Motivated by thermodynamic considerations, we analyse the variation of the quantum mutual information on a unitary orbit of a bipartite system's state, with and without global constraints such as energy conservation.  We solve the full optimisation problem for the smallest system of two qubits, and explore thoroughly the effect of unitary operations on the space of reduced-state spectra.  We then provide applications of these ideas to physical processes within closed quantum systems, such as a generalized collision model approach to thermal equilibrium and a global Maxwell demon playing tricks on local observers. For higher dimensions, the maximization of correlations is relatively straightforward for equal-sized subsystems, however their minimisation displays non-trivial structures. We characterise a set of separable states in which the minimally correlated state resides: a collection of classically correlated states admitting a particular ``Young tableau'' form.  Furthermore, a partial order exists on this set with respect to individual marginal entropies, and the presence of a ``see-saw effect'' for these entropies forces a finer analysis to determine the optimal tableau.
\end{abstract}

\maketitle

\section{\label{intro}Introduction}


The idealisation that a quantum system evolves in a closed environment dates back to the foundations of the theory itself \cite{Schrod}. While this closure preserves the purity of any quantum state, the correlations between its constituent parts can greatly vary. A natural but little-addressed question is how much can these correlations vary when the system undergoes isolated,  unitary dynamics? The issue of varying correlations arises naturally in different thermodynamic processes within closed quantum systems, and can often result in seemingly paradoxical effects.

We address this issue for a bipartite system and use the quantum mutual information (QMI) as a measure of total correlations. The QMI is a quantity that appears within a range of scenarios in quantum thermodynamics, such as work extraction, heat transfer and the approach to equilibrium \cite{FirstPaper}. This paper builds on the work in Ref. \cite{FirstPaper} and illustrates that even the simplest bipartite systems display a rich and non-trivial structure that deserves careful study.

We start in Section \ref{SettingScene} with a declaration of the core mathematical objects and notions that will be used throughout the paper. Section \ref{varcorrU} formally states the key underlying task that we tackle, namely the variation of the quantum mutual information on a unitary orbit. We present the general solution for the maximum QMI in section \ref{secrhomax} and detail what is required for its minimisation in section \ref{rhomin}. Section \ref{sec2qubits} provides a full and explicit solution of this optimisation problem for the instance of two qubits, including the case when a global constraint (e.g. energy conservation) is present. An important tool is introduced: the set $\mathcal{R}$ of eigenvalues for marginals compatible with all states on a unitary orbit of a bipartite quantum system. In section \ref{secAnalysisR} we study this set $\mathcal{R}$ in more detail and observe the effect that unitaries (acting on the global state) have on the marginal spectra.

With these results established, in section \ref{Applications} we describe some applications to physical processes within a closed quantum system. Section \ref{secHeatFlow} discusses the heat flow model of Ref. \cite{FirstPaper} in more detail and covers some important assumptions. In section \ref{secCollision} we establish a more general collision model for equilibration in which the systems retain correlations. Finally, section \ref{secCharlieDemon} illustrates how a global Maxwell-type demon can confuse a local observer by causing an apparent violation of his Second Law.

Section \ref{gendim} presents the minimisation procedure for the QMI in the general dimensional case using a trick that involves enlarging the orbit to its convex hull. This shift of perspective seems to be required in order to prove that the minimum QMI state is classically correlated, and no simpler argument has been found. However the QMI is not constant on this set of classical states, and using majorization techniques we show that the minimum must lie in a subset of these states, most easily described in ``Young tableau'' form.  In section \ref{secSeeSaw} we identify a partial ordering on the representative Young tableaux in terms of the marginal entropies, and discuss a curious geometrical property we call the ``see-saw effect'' which expresses the behaviour of the individual marginal entropies when transforming from one tableau to another, and prohibits us from knowing which tableau is optimal without a deeper analysis, such as that in Ref. \cite{Gary}. Section \ref{secSpecCase} discusses special cases of classical states for which we can unambiguously say how the QMI varies.

\section{\label{SettingScene}Setting the scene}

Let us take a moment to define some of the mathematical objects that will appear throughout.
The global or joint quantum state $\rho$ describing a bipartite system $AB$ is a positive operator on the Hilbert space $\mathcal{H} = \mathcal{H}_A \otimes \mathcal{H}_B$, where $\mathrm{dim}(\mathcal{H})=d$, with unit trace and spectrum $\Lambda = \{\lambda_i\}_{i=1}^d$. It has reduced states $\rho_A$ and $\rho_B$, the states of subsystems $A$ and $B$ obtained by taking the partial traces of $\rho$, which are operators on Hilbert spaces $\mathcal{H}_A$ and $\mathcal{H}_B$. The dimensions of the subspaces are $\mathrm{dim}(\mathcal{H}_A) =d_A$ and $\mathrm{dim}(\mathcal{H}_B) =d_B$ so that $d = d_Ad_B$.

In connection with the physical scenarios we have described in Ref. \cite{FirstPaper}, we use the quantum mutual information (QMI) of the state $\rho$ as a measure of total correlations between its two subsystems $A$ and $B$:
\begin{equation}
\label{QMI}
I(\rho) = S(\rho_A) + S(\rho_B) - S(\rho).
\end{equation}
$ S(\rho)=-\mathrm{Tr}\rho \log \rho = -\sum_{i=1}^d \lambda_i \log \lambda_i$ is the von Neumann entropy of state $\rho$ and the logarithm is taken to base 2. Often we will write $S(\rho)=H(\Lambda)$, $H$ being the Shannon entropy, or for qubits when there are only two eigenvalues $\lambda,1-\lambda$ we use the binary entropy notation $H(\lambda)=-\lambda\log\lambda-(1-\lambda)\log(1-\lambda)$. The QMI is a special case of the relative entropy $I(\rho)=S(\rho||\rho_A \otimes \rho_B) $ and so it is a measure of how distinguishable the state $\rho$ is from the uncorrelated product state of its marginals $\rho_A \otimes \rho_B$. In other words, it measures the total, quantum plus classical, correlations in $\rho$. The relative entropy is non-negative $S(\rho || \sigma) \geq 0$ with equality if and only if $\rho = \sigma$. When $\sigma = \rho_A \otimes \rho_B$ this leads to the subadditivity of entropy $S(\rho_A) + S(\rho_B) \geq S(\rho)$, which ensures that the QMI is non-negative. The QMI is zero if and only if $\rho= \rho_A \otimes \rho_B$, and it is invariant under a local unitary transformation $I(\rho) = I(U_A \otimes U_B \rho U^\dag_A \otimes U^\dag_B)$ or a swap of the states of $A$ and $B$. These symmetry transformations define an equivalence class of states with the same QMI. In the following, when we, for instance, search for a state that gives the lowest QMI, we quote as the solution just one state but it is understood that the full solution is modulo local unitaries and swaps.

When talking about closed evolution, which is essentially a unitary transformation on a system, the concept of a unitary orbit proves useful. Formally this is the set \cite{ModiUO}
$$ \mathcal{O}_{\rho} = \{ U \rho U^{\dag} : \, \forall \, U \in \mathrm{SU}(d) \}$$
and it is called the unitary orbit of $\rho$. Unitary evolution preserves the spectrum of a state, hence $\mathrm{spec}(U \rho U^{\dag}) = \Lambda$.
It follows that the entropy (being a function only of the eigenvalues) of all the states in a unitary orbit is $S(\rho)$.

\section{\label{varcorrU}Extremal correlations attained under unitary transformations}

Our motivating examples in Ref. \cite{FirstPaper} lead to the question of how much a unitary operation can correlate or decorrelate a bipartite state. In other words, we look for $\rho_{min},\rho_{max} \in \mathcal{O}_\rho$ that give the largest and smallest values of  $I(\rho)$.
\newline\newline
Since $S(\rho_{max}) = S(\rho_{min})=H(\Lambda)$, we can write
$$I(\rho_{max}) - I(\rho_{min}) = \Delta S(\rho_A) + \Delta S(\rho_B) := \Delta I$$
where $\Delta S(\rho_\mu) := S(\rho_{max,\mu}) - S(\rho_{min,\mu})$, $\mu = \{A,B\}$ and $\rho_{m,\mu}$ is a reduced state of $\rho_{m}$, $m= \{min,max\}$.

We can not use convex optimisation to solve this because the domain of states is not convex, instead we search for $\rho_{min/max}$ along the unitary orbit separately (later we introduce a trick that allows us to minimise/maximise using convexity properties).
This means determining the reduced states that extremise $S(\rho_{m,A}) + S(\rho_{m,B})$, $m= \{min,max\}$ whilst keeping the spectrum of the composite system fixed.

The entropy function, when it is defined over the spectra of all states in a $d$-dimensional Hilbert space, is a concave function over the set of all possible eigenvalues which, being essentially probability vectors, form a convex set. It ranges from 0 for pure states, whose eigenvalues are $(1,0,0,\dots,0)$ up to a permutation, to $\log d$ for maximally mixed states, whose eigenvalues are uniform and equal to $\frac{1}{d}$.


\subsection{\label{secrhomax}The maximum QMI state $\rho_{max}$}

The derivation for the maximally correlated state on the unitary orbit $\rho_{max}$ appears in Ref. \cite{FirstPaper} and will not be repeated here. The result for $d_A = d_B$ is
\begin{equation}
\label{eqrhomax}
\rho_{max} = \sum_{i=1}^{N} \lambda_i \ket{\Phi_i}\bra{\Phi_i},
\end{equation}
where $\{ \ket{\Phi_i} \}$ is any generalised Bell state basis \cite{GBS}, $N = d_A^2$ and $I(\rho_{max}) = 2\log d_A - H(\Lambda)$. \footnote{When $d_A \neq d_B$ then the maximum value for the QMI is upper bounded by $\log d_A d_B$.}

\subsection{\label{rhomin}The minimum QMI state $\rho_{min}$}

Finding $\rho_{min}$ on the unitary orbit is more difficult than finding $\rho_{max}$ in general. One cannot fully decorrelate an arbitrary state via a unitary transformation hence $I(\rho_{min}) \geq 0$ and, unlike the maximum QMI case, the marginal entropy sum $S(\rho_{min,A}) + S(\rho_{min,B})$ depends on the spectrum $\Lambda$.

The portion of the QMI that varies on a unitary orbit is $S(\rho_A) + S(\rho_B) = H(\Lambda_A) + H(\Lambda_B)$, $\Lambda_\mu = \mathrm{spec}(\rho_\mu)$, hence the domain of interest is the spectra $\Lambda_A, \Lambda_B$ of the reduced states of all the states in $\mathcal{O}_\rho$. Calculating this set of compatible reduced states, given the spectrum of the global state, is rather formidable and is dubbed the ``quantum marginal problem'' \cite{Klyachko,Bravyi,Matthias}.

\subsection{\label{sec2qubits}The primitive case of two qubits}

The simplest possible case is when the subsystems are both qubits, $d_A = d_B = 2$, and we treat this problem separately.

As a physical system it provides the smallest imaginable bipartite scenario, and of course any physical implementation of such a system would require the challenging task of isolating it from the environment. Due to the small size of the systems one would expect proportionally small fluctuations to take place even for moderate temperatures, however as an illustrative example of a bipartite quantum system the two-qubit scenario obeys the same laws as the general $d_A \times d_B$ case, but has the pedagogical advantage of admitting an exact solution.

Let us adopt the convention that the eigenvalues $\lambda_i$ in spectrum $\Lambda$ of the joint state $\rho$ are arranged in non-increasing order $\lambda_1 \geq \lambda_2 \geq \lambda_3 \geq \lambda_4 $ and denote the two eigenvalues of the reduced state $\rho_\mu$ as $\lambda_\mu,\, 1-\lambda_\mu$ where $0\leq \lambda_\mu \leq \frac{1}{2}$, $\mu \in \{ A,B \}$. Then $\rho_A, \rho_B$ are valid reduced states of a state in $\mathcal{O}_\rho$ if and only if their eigenvalues satisfy the following inequalities \cite{Bravyi}
\begin{align}
\label{ineq1}
\lambda_A &\geq \lambda_3 + \lambda_4\\
\label{ineq2}
\lambda_B &\geq \lambda_3 + \lambda_4\\
\label{ineq3}
\lambda_A + \lambda_B &\geq \lambda_2 + \lambda_3 + 2\lambda_4\\
\label{ineq4}
| \lambda_A - \lambda_B | &\leq \operatorname{min}\{ \lambda_1-\lambda_3 , \lambda_2- \lambda_4 \}
\end{align}

\begin{figure}[h!]
\includegraphics[width=3.5in]{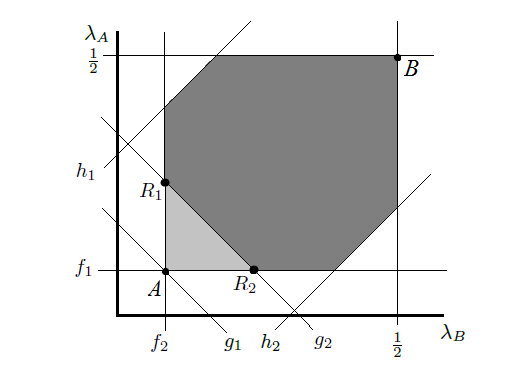}
\caption{Typical set $\mathcal{R}$ of unitary orbit marginal eigenvalues $\lambda_A,\lambda_B$ from inequalities \eqref{ineq1} - \eqref{ineq4}, with equalities equal to the boundary lines labelled $f_p,g_p,h_p$, $p=1,2$. $g_1$ refers to the case when $\lambda_2 = \lambda_3$, light plus dark grey area, and $g_2$ to $\lambda_2 > \lambda_3$, just the dark area (note that the lines $h_p$, $p=1,2$ could be different for these two cases but here they are drawn as the same for simplicity). The marginal eigenvalues for the minimally correlated state $\rho_{min}$ are situated at $A$ when $g_p = g_1$ and at $R_1, R_2$ when $g_p = g_2$. The maximally correlated state $\rho_{max}$ has marginals at $B$.}
\label{figLALBfgh}
\end{figure}

\begin{figure}[h!]
\includegraphics[width=3in]{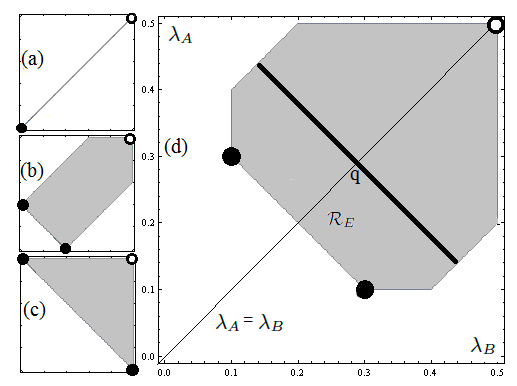}
\caption{Regions $\mathcal{R}$ of allowed  $\lambda_A$ ($y$-axis), $\lambda_B$ ($x$-axis) when the joint state of two qubits, with spectrum $\Lambda = \{\lambda_i\}_{i=1}^4$, has various ranks: $\Lambda =$ (a) \{1,0,0,0\}, (b) \{0.8,0.2,0,0\}, (c) \{0.5,0.5,0,0\}, (d) \{0.6,0.3,0.1,0\} (for the entire shaded region). $\lambda_A, \lambda_B \in [0,\frac{1}{2}]$.
For each spectrum, the filled circles correspond to $\rho_{min}$, the hollow ones to $\rho_{max}$. In (d), a state with energy $E$ defines the set $\mathcal{R}_E$ of states which could have energy $E$. It is bounded from ``above'' by the thick black line, on which the state itself is situated. The maximally correlated state in $\mathcal{R}_E$ is at q.}
\label{figLALBranks}
\end{figure}

A sketch of the region $\mathcal{R}({\Lambda})$ (or just $\mathcal{R}$ when there is no ambiguity) that these inequalities define is presented in figure \ref{figLALBfgh}. The lines $f_p, g_p, h_p$, $p = 1,2$, represent the equality conditions for these inequalities \eqref{ineq1} - \eqref{ineq4}. Figure \ref{figLALBranks} is provided to see the different shapes that $\mathcal{R}$ can have depending on the rank of $\Lambda$ (actually the qualitative difference in the shape of $\mathcal{R}$ depends on the degeneracy of $\Lambda$). In a slight abuse of notation, we use $\mathcal{R}$ for the geometric region of compatible reduced states in $\lambda_A, \lambda_B$ space, the set of coordinates contained within that region and the states $\rho_A, \rho_B$ they correspond to.

Minimising the QMI reduces to finding the minimum of the classical binary entropy sum $H(\lambda_A)+H(\lambda_B)$ over $\mathcal{R}$. It is not quite as simple as it may seem because the domain is not convex, see section \ref{2qubitConvHull}. We first vary each of the terms in the sum separately. Concavity of the entropy (and symmetry about its maximum at $\lambda_{\mu}=\frac{1}{2}$) implies that $H(\lambda_\mu)$ decreases as $\lambda_\mu$ tends to zero. The line that bounds $\mathcal{R}$ closest to zero is $g_p$: $\lambda_A + \lambda_B = \lambda_2 + \lambda_3 + 2\lambda_4$ so the minimum of $H(\lambda_A)+H(\lambda_B)$ occurs somewhere on this boundary of $\mathcal{R}$. We show in Appendix A that the minimum of this function occurs at points $R_1, R_2$ in figure \ref{figLALBfgh} where
\begin{align}
\lambda_A = \lambda_3 + \lambda_4\\
\lambda_B = \lambda_2 + \lambda_4
\end{align}
or vice versa,
and the joint state (modulo the equivalence class of the QMI) which corresponds to it is
\begin{align}
\label{eqrhomin0}
\rho_{min} = \sum_{j,k=1}^0 \lambda_{jk} \ket{e_j}\bra{e_j} \otimes \ket{f_k}\bra{f_k}
\end{align}
where $\lambda_{jk}$ is a reindexing of $\lambda_i$ ($\lambda_{00} = \lambda_1, \lambda_{01} = \lambda_2, \dots$) and $\{\ket{e_j}\},\{\ket{f_k}\}$ are qubit basis states for systems $A$ and $B$ respectively.
The QMI for two qubits has a minimum value of
$$I(\rho_{min}) = H(\lambda_3 + \lambda_4) + H(\lambda_2 + \lambda_4) - H(\Lambda). $$
Here we have used the binary entropy notation: $H(\lambda_3 + \lambda_4) \equiv H(\lambda_1 + \lambda_2) \equiv H(\lambda_1 + \lambda_2,\lambda_3 + \lambda_4)$.

Any state $\rho = \sum_{i=1}^4 \lambda_i \ket{\psi_i}\bra{\psi_i}$ diagonal in the basis ${\ket{\psi_i}}$ along the unitary orbit of $\rho_{min}$ can be transformed to $\rho_{min}$ by the unitary $U_{min} = \sum_{i=1}^4 \ket{ef_i}\bra{\psi_i}$ where $\ket{ef_i} = \ket{e_j}\ket{f_k}$ are products of qubit basis states and $j,k = 0,1$ are the binary representation of $i$.

So to answer the initial question, for a two-qubit system, the maximum that the QMI can change by over the domain of states in $\mathcal{O}_\rho$ with  $\mathrm{spec}(\rho) = \{\lambda_i\}_{i=1}^4$, $\lambda_i \geq \lambda_{i+1}$, is
\begin{align}
\label{deltaI}\Delta I = 2 - H(\lambda_1+\lambda_2) - H(\lambda_1+\lambda_3).
\end{align}

\subsection{\label{secEcons}The Imposition of a Global Constraint: Energy Conservation}

In this section we seek the maximum change in the QMI 
\begin{equation}
\Delta I_{E} = I(\rho') - I(\rho) = S(\rho'_A) + S(\rho'_B) - S(\rho_A) - S(\rho_B)
\end{equation}
subject to $\rho' = U \rho U^{\dag}$ where $U$ is now an energy-conserving unitary satisfying $E=\Tr(\rho H) = \Tr(\rho' H)$. $H$ is a Hamiltonian for the composite system $AB$, taken to be local $H = H_A + H_B$, and the initial reduced states are thermal with respect to $H_A, H_B$. These are the assumptions for the heat flow model between two quantum systems, Example 2 in Ref. \cite{FirstPaper}. Below we find the maximum $\Delta I_{E}$ when A and B are qubits.


We take the local Hamiltonians $H_A, H_B$ of the initial subsystem states to be equal \footnote{The following results also hold when $H_A = U H_B U^{\dag}$ but not if the energy spacings between the ground and excited levels differ. The reason (c.f. Appendix B) is that this defines an energy conserving region that is bounded by a skewed line, it is of the form $a \lambda_A + b \lambda_B = C$, where $a \neq b$ and $C$ is a constant, and the maximally correlated state is no longer at the point where $\lambda_A = \lambda_B$.}: the ground and excited states energies are zero and one respectively and the energy eigenbasis is the computational basis $\{ \ket{0},\ket{1}\}$ thus $H_\mu = \ket{1}\bra{1}_\mu$. We can write the initial (thermal) reduced state of $\rho$ as
$$\rho_\mu = (1-\lambda^i_\mu)\ket{0}\bra{0} + \lambda^i_\mu\ket{1}\bra{1},$$
where $\lambda^i_\mu=e^{-1/T_\mu}/\Tr(\rho_\mu) \leq \frac{1}{2}$ and $T_\mu$ is the temperature of system $\mu$.

Let us suppose that during the energy exchange interaction (a global energy conserving unitary on $\rho$) correlations decrease, i.e. the two systems $A, B$, though locally thermal, are initially correlated. 
We would therefore like to minimise $S(\rho'_A) + S(\rho'_B)$ subject to total average energy conservation. The initial total energy set by the initial state $\rho$ is
\begin{equation}
E  = \Tr(\rho H) = \Tr(\rho_A H_A) + \Tr(\rho_B H_B) = \lambda^i_A + \lambda^i_B
\end{equation}
with $0 \leq E \leq 1$.

\textit{Note}: The point $(\lambda^i_A, \lambda^i_B) \in \mathcal{R}$ is on the line $\lambda_A + \lambda_B = E$, the thick black line in figure \ref{figLALBranks} (d). The joint states represented by this line all have total energy $E$ and their reduced states are thermal. The QMI, or equivalently just $H(\lambda_A) + H(\lambda_B)$, is concave on this line, which is easily seen when the function is written as $H(\lambda_A) + H(E-\lambda_A)$, and its maximum occurs when $\lambda_A = \lambda_B = \frac{E}{2}$, point q in the figure. Hence the maximum possible initial value of the QMI for a state with total average $E$ is $2 H(\frac{E}{2})$.

After the global unitary transformation $U \rho U^{\dag}  =\rho'$, the local subsystem states evolve to $\rho'_\mu = (1-\lambda'_\mu)\ket{v}_\mu\bra{v}_\mu + \lambda'_\mu\ket{\bar{v}}_\mu\bra{\bar{v}}_\mu$ with $\lambda'_\mu \leq \frac{1}{2}$ and $\{\ket{v}_\mu,\ket{\bar{v}}_\mu \}$ is a qubit basis
\begin{align}
\ket{v}_\mu &= \cos \frac{\theta_\mu}{2} \ket{0} - e^{-i\phi_\mu} \sin \frac{\theta_\mu}{2} \ket{1} \\
\ket{\bar{v}}_\mu &= \sin \frac{\theta_\mu}{2} \ket{0} + e^{i\phi_\mu} \cos \frac{\theta_\mu}{2} \ket{1}
\end{align}
The final total average energy $E'= \Tr(\rho'_A H_A) + \Tr(\rho'_B H_B)$ ($\rho_A, \rho_B$ are not constrained to being diagonal in $H_A, H_B$) is
\begin{equation}
\label{EconsThL}
E'  = 1 - \frac{1}{2}(\cos \theta_A + \cos \theta_B) + \lambda'_A \cos \theta_A +\lambda'_B \cos \theta_B
\end{equation}
In order for energy to be conserved, $E' = E$. This condition and equation (\ref{EconsThL}) gives a collection of straight lines, defining a region  $\mathcal{R}_E \subseteq \mathcal{R}$ representing joint states which could have total energy $E$ for some choice of angles $\theta_A,\theta_B$.

$\mathcal{R}_E$ is given by $\lambda_A + \lambda_B \leq E$, for the derivation see Appendix B and for a visualisation see figure \ref{figLALBranks} (d). The energy conserving region always contains the minimally correlated state, hence the greatest possible change in the QMI at constant average energy $E$ occurs when the initial state occurs when $\lambda^i_A = \lambda^i_B = \frac{E}{2}$ (these marginals could define multiple global states, even after factoring out the states in the equivalence class of the QMI, see section \ref{MultiplePoints}) and the final state is the minimally correlated one, $\rho_{min}$ given by equation \eqref{eqrhomin0}, with $\theta_A, \theta_B$ chosen such that equation \eqref{EconsThL} is satisfied (there is a range of angles which do this). We have then that
\begin{align}
\label{deltaIheatres}
\Delta I_{E} = 2H\left(\frac{E}{2}\right) - H(\lambda_1 + \lambda_2) - H(\lambda_1 + \lambda_3)
\end{align}

The way to think about the region $\mathcal{R}_E$ intuitively is that it contains final states whose marginals in general have longer Bloch vectors than the initial $\rho$, achieving this higher local purity is made possible by using up the correlations in the joint state. The constant energy condition is met by rotating the Bloch vectors off the energy axis by the angles $\theta_A, \theta_B$, thereby adding coherence what was originally a thermal state. 

Note that if we insist the also the final states are thermal, then the energy conserving region is just the set of states in $\mathcal{R}$ that are on the line $\lambda_A + \lambda_B = E$. The maximally correlated state is then still at $\lambda_A = \lambda_B = \frac{E}{2}$ but the minimally correlated one is at the edge of the line.

\section{\label{secAnalysisR}Analysis of the set of reduced state spectra $\mathcal{R}$}

We appealed to the collection $\mathcal{R}$ of reduced states compatible with a joint state having a fix spectrum $\Lambda$ in order to find the maximally and minimally correlated states, with and without an energy conservation constraint. It would be useful to know how the action of a global unitary acting on the joint state translates to dynamics in $\mathcal{R}$, but first we must tackle an issue of degeneracy.

\subsection{\label{MultiplePoints}One point refers to a range of compatible joint states}

A two qubit density matrix is specified by 15 parameters, a point in $\mathcal{R}$ defines only 5 independent parameters $(\lambda_A, \lambda_B, \lambda_1, \lambda_2, \lambda_3)$, hence there is a lot of degeneracy as each point in $\mathcal{R}$ corresponds to a whole set of compatible joint states. Modulo the parameters that don't affect the QMI (the directions of the Bloch vectors, which requires 4 more parameters) there are still 6 parameters that can be freely chosen. In appendix A we show that the coordinate in $\mathcal{R}$ that represents $\rho_{min}$ is given by a unique state (modulo local unitaries and swaps). However there are points in $\mathcal{R}$ for which this is not true, these points determine states that have the same QMI but are not related by local unitaries or swaps, hence they would have different amounts of correlation, or entanglement, if we used another measure that, say, depended on the correlation- or T-matrix of the state, whose elements are $t_{ij} = \Tr(\sigma^i \otimes \sigma^j \rho )$, $i,j = 1,2,3$ and $\sigma^i$ is a Pauli matrix. We illustrate this with an example of a ``triple point'' in $\mathcal{R}(\Lambda)$. Defining
\begin{widetext}
\begin{align}
\label{rhothetaphi}
\rho(\alpha,\beta,\gamma,\delta, \cos\theta, \cos\phi) :=
\frac{1}{2}\left(
  \begin{array}{cccc}
    \alpha+\delta+(\alpha-\delta)\cos\phi & 0 & 0 & (\delta-\alpha)\sin\phi \\
    0 & \beta+\gamma + (\beta-\gamma)\cos\theta & (\gamma-\beta)\sin\theta & 0 \\
    0 & (\gamma-\beta)\sin\theta & \beta+\gamma -(\beta-\gamma)\cos\theta  & 0 \\
    (\delta-\alpha)\sin\phi & 0 & 0 & \alpha+\delta-(\alpha-\delta)\cos\phi \\
  \end{array}
\right),
\end{align}
\end{widetext}
where the matrix is written in the computational basis,
three states that correspond to the same $\lambda_A, \lambda_B$ but are not connected by a symmetry of the QMI are $\rho(\lambda_1,\lambda_2,\lambda_3,\lambda_4,\frac{1}{\sqrt{2}},\frac{\lambda_3-\lambda_4}{\lambda_1 - \lambda_4})$, $\rho(\lambda_2,\lambda_1,\lambda_3,\lambda_4,0,\frac{\lambda_3-\lambda_4}{\lambda_2 - \lambda_4})$,
$\rho(\lambda_3,\lambda_1,\lambda_2,\lambda_4,0,1)$ and

\begin{align}
\lambda_A = \lambda_B = \frac{1}{2}(1 - (\lambda_3 - \lambda_4)).
\end{align}

We know also of ``double points'' and correspondingly $\rho_{min}$ is a ``single point'', shown in Appendix A.

Despite the value of the QMI not being sensitive to these different states, their dynamics under energy conserving unitaries leads to different values of $\Delta I$, and it is precisely the change in QMI which is the physically relevant quantity, see Ref. \cite{FirstPaper}: it is in this way that these hidden correlations are revealed.

\subsection{Action of unitaries on $\rho$}

There are many ways in which we could analyse the effect of a global unitary on a two qubit state, with regards to its position in $\mathcal{R}$. To simplify the task, we concentrate on some ``elementary'' transformations only rotating in two-dimensional subspaces and discuss what their consequences are for energy conservation.

Let us assume that the local bases are fixed in the computational basis, then the individual and total Hamiltonians are $H_A = H_B = \ket{1}\bra{1}, H = H_A \otimes \mathbb{I} + \mathbb{I} \otimes H_B = \ket{01}\bra{01} + \ket{10}\bra{10} + 2\ket{11}\bra{11}$. There are two key unitary transformations (matrices are all written in the computational basis) that are effective two-dimensional rotations:
\begin{eqnarray}
\label{Uodd}
U^{(o)}(\theta) &=& \exp(-iH^{(o)}\theta/2)\\
&=&\left(
  \begin{array}{cccc}
    1 & 0 & 0 & 0 \\
    0 & \cos(\frac{\theta}{2}) & \sin(\frac{\theta}{2}) & 0 \\
    0 & -\sin(\frac{\theta}{2}) & \cos(\frac{\theta}{2}) & 0 \\
    0 & 0 & 0 & 1 \\
  \end{array}
\right) \\
\label{Ueven}
U^{(e)}(\phi) &=& \exp(-iH^{(e)}\phi/2)\\
&=&\left(
  \begin{array}{cccc}
    \cos(\frac{\phi}{2}) & 0 & 0 &  \sin(\frac{\phi}{2}) \\
    0 & 1 & 0 & 0 \\
    0 & 0 & 1 & 0 \\
    -\sin(\frac{\phi}{2}) & 0 & 0 & \cos(\frac{\phi}{2}) \\
  \end{array}
\right)
\end{eqnarray}
where the interaction Hamiltonians are $H^{(o)} = \frac{1}{2}(X\otimes Y - Y\otimes X)$ and $H^{(e)} = -\frac{1}{2}(X\otimes Y + Y\otimes X)$, $X, Y$ are Pauli matrices. Since $[H^{(o)}, H] = 0$, $U^{(o)}$ is a \textit{strong energy conserving unitary} (SECU). In terms of its action on the reduced state eigenvalues (and therefore on the QMI when starting from states that are locally diagonal in $H_A, H_B$ which is all that we require, see example 2 in Ref. \cite{FirstPaper}) it is the most general unitary that commutes with the total Hamiltonian.

The matrix $U^{(o)}(\theta)$ rotates in the energy degenerate ``odd'' subspace $\{\ket{01}, \ket{10}\}$ and $U^{(e)}(\phi)$ rotates in the ``even'' subspace $\{\ket{00}, \ket{11}\}$, but is not energy conserving. $U^{(o)}(\theta)$ and $U^{(e)}(\phi)$ commute. Let the initial state be diagonal in the computational basis
\begin{align}
\label{rho0}
\rho_0 =
\left(
  \begin{array}{cccc}
    \alpha & 0 & 0 &  0 \\
    0 & \beta & 0 & 0 \\
    0 & 0 & \gamma & 0 \\
    0 & 0 & 0 & \delta \\
  \end{array}
\right)
\end{align}
and the elements on the diagonal are some permutation of eigenvalues $\Lambda = \{\lambda_i\}_{i=1}^4$. To see how the unitaries transform $\rho_0$ in the region $\mathcal{R}$ we need to look at the (lowest) eigenvalue of the reduced states. A state evolved under these unitaries looks like $\rho(\alpha,\beta,\gamma,\delta, \cos\theta, \cos\phi)$ in equation (\ref{rhothetaphi}), it is often called an ``X-state''. Note that $\rho(\alpha,\beta,\gamma,\delta, \cos\theta, \cos\phi)$ has reduced states that are diagonal in the local bases $H_A, H_B$, that is, its local states are thermal (the spectrum can be equated to a Gibbs distribution). In this notation $\rho_0 = \rho(\alpha,\beta,\gamma,\delta, 1, 1)$.

The reduced state eigenvalues of $\rho(\alpha,\beta,\gamma,\delta, \cos\theta, \cos\phi)$ are
\begin{eqnarray}
\label{genLA}
\lambda_A(\alpha,\beta,\gamma,\delta, \cos\theta, \cos\phi) &=&\\
 \frac{1}{2}\left( 1 - (\beta - \gamma)\cos\theta - (\alpha - \delta)\cos\phi \right)\nonumber \\
\label{genLB}
\lambda_B(\alpha,\beta,\gamma,\delta, \cos\theta, \cos\phi) &=&\\
 \frac{1}{2}\left( 1 + (\beta - \gamma)\cos\theta - (\alpha - \delta)\cos\phi\right)\nonumber
\end{eqnarray}
if we pick $\min\{\alpha - \delta,\beta - \gamma\} = \beta - \gamma$. Care must be taken to ensure that these eigenvalues are the smallest depending on the ordering used for $\alpha,\beta,\gamma,\delta$.

Following from this, $\lambda_A + \lambda_B = 1 - (\alpha - \delta)\cos\phi$, hence varying $\phi$ provides a set of lines of the form $\lambda_A + \lambda_B = C_1$. On the other hand, $\lambda_B - \lambda_A = (\beta - \gamma)\cos\theta$, thus $\theta$ defines the lines $\lambda_A = \lambda_B + C_2$. Therefore, when the initial state is that given in equation (\ref{rho0}),  we can describe the evolution of the odd and even unitaries fairly simply in terms of the marginals and the regions that they trace out in $\mathcal{R}$. It is easy also to see their effect on the QMI (in simplified notation):
\begin{align}
I(\cos\theta = x) = H(c - m x) +  H(c + m x) - H(\Lambda)\\
I(\cos\phi = y) = H(a - n y) +  H(b - n y) - H(\Lambda)
\end{align}
where $a,b,c,m,n$ are all constants related to terms in equations (\ref{genLA}) and (\ref{genLB}), and
the QMI is (separately) concave over the domains $x$ and $y$.

\begin{figure}[h!]
\includegraphics[width=3.5in]{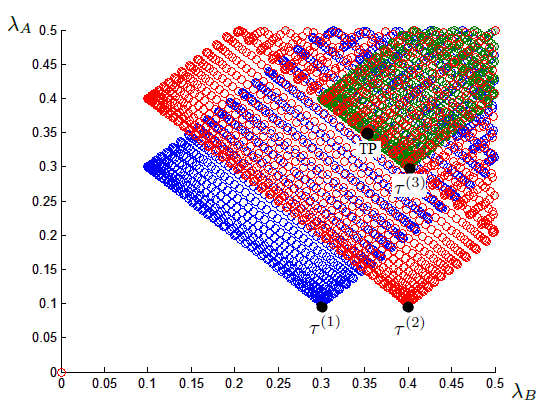}
\caption{(Color online) The reduced state eigenvalues of $U^{(o)}(\theta) \rho_0 U^{(o)}(\theta)^{\dag}$,  $U^{(e)}(\phi) \rho_0 U^{(e)}(\phi)^{\dag}$, $0 \leq \theta, \phi \leq \pi$ when $\rho_0 = \tau^{(i)}$, $i=1$, in blue, $i=2$, in red and $i=3$ in green. $\tau^{(i)}$ is a classically correlated state whose marginals are also plotted and a triple point TP is shown when three states with different T-matrices but same values of QMI coincide. The global spectrum is $\lambda_1 = 0.6, \lambda_2 = 0.3, \lambda_3 = 0.1, \lambda_4 = 0$.}
\label{figUeUofill}
\end{figure}

It is shown in section \ref{MinYT2qubit} that of the $4!$ possible orderings of $\lambda_i$, $i=1,2,3,4$, there are only three classes of permutations which give distinct values of the QMI. Representative states for each class are taken to be $\tau^{(1)} = \rho(\lambda_1,\lambda_2,\lambda_3,\lambda_4, 1, 1)$, $\tau^{(2)} = \rho(\lambda_2,\lambda_1,\lambda_3,\lambda_4, 1, 1)$, $\tau^{(3)} = \rho(\lambda_3,\lambda_1,\lambda_2,\lambda_4, 1, 1)$. Note again that, especially when calculating energies, care must be taken so that the reduced states of each of these are in the form $\rho_\mu = (1-\lambda_\mu)\ket{0}\bra{0} + \lambda_\mu\ket{1}\bra{1}$ with $0  \leq \lambda_{\mu}  \leq \frac{1}{2}$ - local unitaries may be used to ensure this condition. Using each of these as an initial state $\rho_0$, figure \ref{figUeUofill} depicts the effect of $U^{(e)}, U^{(o)}$ on the each of their marginals.

We can clearly see the triple point described in section \ref{MultiplePoints}, labelled TP in figure \ref{figUeUofill}, as well as many other overlaps corresponding to triple and double points. A corollary of this is that starting at a triple/double point, the unitary $U^{(o)}$ generates portions of the line $\lambda_A + \lambda_B = C_1$ that are of varying length, depending on which exact state we started with. The longer the length, the lower the correlations can go since the QMI is concave on the line.

\begin{figure}[h!]
\includegraphics[width=3.5in]{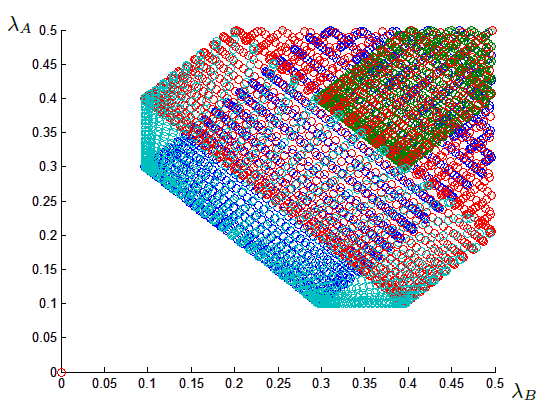}
\caption{(Color online) Same as figure \ref{figUeUofill} with the marginal eigenvalues of $\tilde{U}(\xi) \tau^{(1)}\tilde{U}(\xi)^{\dag}$ added, $0 \leq \xi \leq \pi$, this now is equal to $\mathcal{R}$.}
\label{figUoUeUtFill}
\end{figure}

Comparing figure \ref{figUeUofill} with the full allowed region $\mathcal{R}(\Lambda)$, see for instance figure \ref{figLALBranks} (d), we see that just these two unitaries can evolve the classically correlated state to almost every point in $\mathcal{R}(\Lambda)$.  The ``wings'' that are missing in figure \ref{figUeUofill} can be obtained using one more unitary
\begin{eqnarray}
\tilde{U}(\xi) &=& \exp(-i\tilde{H}\xi/2)\nonumber\\
&=&\left(
  \begin{array}{cccc}
    \cos(\frac{\xi}{2}) & \sin(\frac{\xi}{2}) & 0 &  0 \\
    -\sin(\frac{\xi}{2}) & \cos(\frac{\xi}{2}) & 0 & 0 \\
    0 & 0 & 1 & 0 \\
    0 & 0 & 0 & 1 \\
  \end{array}
\right)
\end{eqnarray}
with $\tilde{H} = -\frac{1}{2}(\mathbb{I}\otimes Y + Z \otimes Y)$.
This unitary acting on $\rho(\lambda_1,\lambda_2,\lambda_3,\lambda_4, \cos\theta, 1)$ results in figure $\ref{figUoUeUtFill}$. By considering the transformation this unitary generates on the marginal eigenvalues, it is simple to show that these three unitaries generate all points in $\mathcal{R}$. To summarise, we have identified how to reach all points in the allowed region $\mathcal{R}$ starting from classically correlated states using just three unitaries, overlaps are easily seen to be double/triple points and the joint state they correspond to is known.

\bigskip

Let us take a closer look at unitary evolution for states in the energy conserving region $\mathcal{R}_E \subseteq \mathcal{R}$, defined as the set fulfilling $\lambda_A + \lambda_B \leq E$. We set the initial state now as the maximally correlated state for this subregion $\rho_0 = \rho_E^{max}$, it is at the point on the line $\lambda_A + \lambda_B = E$ where $\lambda_A = \lambda_B$. The local and total Hamiltonians are still the same (diagonal in the computational basis). To conserve energy, the unitary evolution of $\rho_0$ must obey
$$ E = \Tr(\rho_0H) = \Tr(U \rho_0 U^{\dag} H)$$
There are two types of unitary that satisfy this: SECUs $U_{S}$ that commute with the total Hamiltonian (above we stated that the only SECU for this Hamiltonian is $U^{(o)}$) and weak energy conserving unitaries (WECUs) $U_{W}$ that do not commute with the total Hamiltonian but preserve the total energy on average.

Depending on the exact form of $\rho_E^{max}$ (could be e.g. a triple point) the SECU evolves it along some portion of the line $\lambda_A + \lambda_B = E$.
Getting to other states in the region $\mathcal{R}_E$, for instance where $\lambda_A + \lambda_B < E$, and conforming with the energy conserving condition, equation (\ref{EconsThL}), requires WECUs, which can shrink $\lambda_A, \lambda_B$, and local unitaries to rotate the local bases by appropriate angles $\theta_A, \theta_B$. This generally amounts to elongating the local Bloch vectors and rotating them off the energy axis in order to preserve total average energy. Consequently, the reduced states are permitted to be non-diagonal in the energy eigenbasis and so they are no longer thermal, however we argue in Ref. \cite{FirstPaper} and the next section that this is still physically and thermodynamically relevant. Note there is a range of angles $\theta_A, \theta_B$ that satisfy equation (\ref{EconsThL}) and the amount energy exchanged between the qubits depends on these angles.

There is an interesting conclusion to be noted from all of this: sometimes a SECU can only transform a state with total energy $E$ along a portion of the line $\lambda_A + \lambda_B = E$ but there are states that are on $\lambda_A + \lambda_B = E$ in $\mathcal{R}$ that are not reachable by the SECU \textit{and} these unreachable states are locally thermal. This means that a WECU is required to get to these states. Hence if we, say, restricted our heat flow model (in the next section) to initial \textit{and} final states being thermal, with total energy conserved, it would not be general enough to only consider SECUs, one would still have to include the WECUs for the full analysis.

Another property of transformations under unitaries that commute with the Hamiltonian is that the quantum variance $\delta = \Tr[\rho H^2] - (\Tr[\rho H])^2$ remains constant as $\rho \rightarrow U_S(t) \rho U_S(t)^{\dag}$. We assume constant average energy $\Tr[\rho H] = E$ so changes in $\delta$ only depend on the first term. All moments of $H$, $\Tr[\rho H^k]$ for $k = 1,2...$ are constant for SECUs. This means that the statistics of all the measurements of the energy of $U_S(t) \rho U_S(t)^{\dag}$ will be the same, due to total constant energy and the reduced states being diagonal in the energy eigenbasis. Quantum uncertainty due to coherence is introduced when the observable and the state are not locally codiagonal, and in general the term $\Tr[\rho H^2]$ will increase for unitaries the do not commute with $H$. 

\section{\label{Applications}Applications to physical processes within a closed quantum system}

Having developed a body of results in the previous section we apply this work to several scenarios that arise in quantum thermodynamics \cite{FirstPaper}, such as equilibration and heat flow between two initially thermal states. As before, the underlying principles and equations are quite general,  and for simplicity we make extensive use of the illustrative two-qubit case, and highlight points where it differs qualitatively from the more general case.

\subsection{\label{secHeatFlow}Heat Flow}

The model for the heat flow, example 2 in Ref. \cite{FirstPaper}, between two systems $A$ and $B$ is based on three assumptions: (1) the global evolution of the composite system $\rho$ is unitary $\rho \rightarrow \rho' = U \rho U^{\dag}$, (2) the total average energy of $\rho$ is conserved, $E=\Tr(\rho H) = \Tr(\rho' H)$ \footnote{The total Hamiltonian is local $H = H_A + H_B$ and $H_A, H_B$ are Hamiltonians for the reduced states of $\rho$ but not necessarily of $\rho'$.}
and (3) the reduced states $\rho_A, \rho_B$ of the initial state $\rho$ are thermal.


This model places a limit on the amount of heat that can flow from one system to another. It adheres to conditions (1)-(3) above and, using the fact that the free energy is minimised by a thermal state, yields the inequality \cite{FirstPaper}
\begin{align}
\label{QTI} Q_A \left(\frac{1}{kT_A} - \frac{1}{kT_B}\right) \geq \Delta I_{\mathrm{heat}},
\end{align}
where $Q_A$ is the heat into system A, $T_A$ $(T_B)$ is the initial temperature of system A (B), $k$ is the Boltzmann constant and $\Delta I_{\mathrm{heat}}$ is the difference in the mutual information before and after the heat exchange interaction, its greatest value for two qubits is displayed in equation \eqref{deltaIheatres}, $|\Delta I_{\mathrm{heat}}| \leq \Delta I_E$.

\subsection{\label{secPhysDigress}A Digression on Physical Assumptions}

To reduce misconceptions, we would like to discuss several key points about the physical assumptions underlying the application of our formal results to physical scenarios involving closed quantum systems. We use the heat flow model as a base for our discussion, but bear in mind that it is not limited to this one situation, nor is it restricted to qubit systems.

\subsubsection{Range of Applicability}

The first issue that one may raise about inequality \eqref{QTI} is on the usage of the free energy and the definition of temperature. Only the initial temperatures of the systems A, B appear in \eqref{QTI}, and the temperatures of the final states may be different or may not even be defined (they are not restricted to be thermal states). If this is so, how can the free energy be used here? The answer is that we employ the \textit{free energy functional} $F_{H,T}(\rho)$ which is initially an information theoretic quantity and, as a mathematical object defined over all state space, can be applied to any quantum state, not just Gibbs states.  The Gibbs, or thermal, states $\rho^{th} = \exp(- H / kT)/Z$ are special in as much as they are the states that minimise the free energy functional $$F_{H,T}(\rho) = \Tr[H \rho] - TS(\rho),$$ $S(\rho)$ is the von Neumann entropy of $\rho$. $F_{H,T}(\rho^{th})$ coincides with the usual thermodynamic free energy result. In fact a change in the free energy functional $F_{H,T}(\rho) - F_{H,T}(\rho^{th})$ is proportional to the relative entropy $S(\rho || \rho^{th})$, which has an operational meaning in terms of the distinguishability of two states, and emphasises the physical relevance of the free energy functional. The positivity of the relative entropy for all quantum states in turn corresponds to a directionality for any thermodynamic process: $\Delta F_{H,T} \geq 0$. In standard situations this corresponds to the free energy being minimised by an equilibrium state. The heat flow goal is then to explore the asymmetry in hot $\rightarrow$ cold and cold $\rightarrow$ hot heat flow from Eq. \eqref{QTI}, which is a broad technical constraint dependent only on initial temperatures and does not rely on defining final temperatures.

\subsubsection{Local Thermality and Entanglement}

The free energy functional constrains any changes in local variables, irrespective of coupling strength, and it is the physically relevant quantity that provides the directionality for such internal statistical processes. Local concepts, such as the initial temperatures of A and B, can still limit non-local properties: the reduced states are initially entirely indistinguishable from thermal states (when restricted to doing only local operations on them) but the presence of correlations means that we do not assume ``molecular chaos" \cite{Partovi}.  When one works in this broader framework we still have directionality for any energy conserving process which comes from the positivity of the relative entropy. The dependence on temperatures just reflects the assumption of initial thermal marginals but also places a limit on the amount of correlations possible (for example two systems which are locally ``too cool" cannot be maximally entangled), which in turn bounds any anomalous heat flow.

It is very much permissible to have a bipartite system whose marginals are thermal but the global system is allowed to be correlated or even entangled. Indeed recently it has been shown \cite{PopescuShortWinter} that local thermality arises typically for constant energy pure states. The degree of correlation is also something we, a priori, do not limit, however any correlation in $\rho$ must be consistent with its reduced states $\rho_A$, $\rho_B$ being (initially) Gibbs states. This condition means that outcomes of local operations on, say, $\rho_A$ must be indistinguishable from doing the same operations on an uncorrelated thermal state with the same temperature and Hamiltonian. The correlations amount to extending the description of the system from local observables $\{H_A, H_B\}$ to $\{H_A, H_B, H_A \otimes H_B\}$.

\subsubsection{Bipartite Interactions and Heat Flow}

Let us clarify what it means when we say systems A and B interact. Interacting here means that the systems undergo a global unitary evolution, which conserves the total energy. This evolution could be thought of in different ways. It could be controlled by an experimenter who switches on an interaction $V(t)$ for some finite time. Or we can think of it as a scattering process: initially the particles are assumed to not interact, e.g. they are far apart, hence $H_i = H_A + H_B$, but there could already exist an unknown degree of correlation between them. The role of the experimenter would then be to bring the particles together so the particles can interact via some natural entropy and energy conserving process. The only action the experimenter takes is in bringing the particles together and then apart, controlling the interaction strength (how close they are) and the interaction time. Both these circumstances are modelled as: Total Hamiltonian $H = H_A + H_B$ for $t < 0$, $t > T$, and $H = H_A + H_B + V(t)$ for $0 \leq t \leq T$, and no net work is done on the total system by assumption. Note there is no explicit time $t$ in our model but we assume there is some ``clock" that governs the temporal length of the interaction.

During part of the total interaction time $T$, the experimenter may have to put in or extract energy out of the system in order to controllably interact A and B, however the total system AB is required to finish in a state such its total energy is equal to its initial total energy, and so overall the experimenter does no work. In this way we can formulate energy conservation in our heat exchange model as a zero work process: $W = \Tr[H \rho^f_{AB}] - \Tr[H \rho^i_{AB}] = 0$, $H = H_A + H_B$. Energies exchanged between the subsystems are assumed inaccessible for external work. As a result we call the internal exchanged energy ``heat". This is in turn supported by the fact that the local entropies change. This is still a valid measure of average energy even when, say for system A, $[\rho^f_A, H_A] \neq 0$, because energies are measured with respect to the original energy eigenbasis defined by $H_A$ even when the final states are no longer diagonal in this basis. This is validly done elsewhere in the literature, see for instance Refs. \cite{QMaxWork} and \cite{ResourceTD2}.




Normally heat flowing from cold to hot is associated with an external source of energy or heat pump driving the process, however this can also occur within a closed system by using up internal correlations. In this sense, correlations are a resource. The model is in a way an embodiment of the Clausius form of the second law: ``No process is possible whose sole result is the transfer of heat from a body of lower temperature to a body of higher temperature ".




One usually considers thermodynamic processes from an ``outside view" where heat baths and work reservoirs externally control the system. However one can switch to an ``inside view", where external influences are described as effective changes in the system's Hamiltonian. This leads to an alternative ``inside view" definition of heat and work \cite{DiffHeatWork}. The unconventionality in our heat exchange scenario arises from the fact that we do not assume interactions with heat baths or work reservoirs, all of our interest lies in the internal view in order to obtain as general a  set of constraints on the evolution as possible - i.e with as little model dependence as possible. We restrict ourselves to a local thermodynamic level of description, the presence of correlations manifests itself in negative heat flow, inferred from changes in $\langle H_A \rangle$, $\langle H_B \rangle$.  The actual unitary that governs the interaction of the systems A and B may be unknown to the observer. Thus the internal view in \cite{DiffHeatWork}
could in principle be applied here, but this is not our regime of interest, since the observer would have to know the initial correlations between A and B and the type of unitary used.

Stationary heat flow is normally maintained if each system is coupled to a large external heat bath leading to dissipation and irreversibility. If this were the case then the effect of correlations might be washed out. However we work in a different regime: we do not want to assume an open system, there is no environment, no dissipation and we focus on internal processes of closed systems. Energy exchange occurs only between A and B (which in principle can be any size) and it is controlled by the interaction ``time".  Once A and B stop interacting, they are static and energy exchange desists. It is in this way that we obtain heat exchange - not by dissipation. Since A and B have relaxed after the interaction, they can be assumed to be ``in equilibrium" (but not necessarily Gibbs states) since their states may (see below) remain stationary.

\subsubsection{Strong versus Weak Coupling}

We now turn to the more subtle issue of coupling strength. It has been shown that without weak coupling, serious problems arise \cite{WeakStrCoup1,WeakStrCoup2} such as apparent local violations of Landauer's principle. We make no assumptions about coupling strength in our model, nor do we need to. First we must determine what exactly is meant by coupling strength in our situation; in interacting systems there can be a notion of internal and external coupling strengths. Usually coupling strength is a term associated with open systems, i.e. the interaction of a system with a thermal bath, the total Hamiltonian of the system being $H = H_A + H_B + g H_{int}(t)$, so we can make the identification $V(t)= gH_{int}(t)$, where $g$ is the coupling strength. Roughly speaking, in an open system approach, A might be low dimensional and B a bath (infinite dimensional). When $g$ is small (weak coupling) the subsystem of interest, A, is assumed to stay diagonal in its original local energy eigenbasis, but in the long time limit thermalises to a Gibbs state having the temperature of the bath. When $g$ is large (strong coupling) the joint system AB evolves to a Gibbs state, with Hamiltonian $H$, and A is no longer diagonal in $H_A$.

As mentioned, we do not have an interaction ``always on" but consider a transitory interaction (scattering). Before and after, $H = H_A + H_B$ and no ambiguity arises in the specification of subsystems and their individual properties. As such we need not impose restrictions on the transitory $V(t)$. A useful comparison might be in the formulation of thermodynamics by Lieb \& Yngvason \cite{LiebYng}. In formulating the principle of entropy increase from the second law, Lieb \& Yngvason define a thermodynamic process via ``adiabatic accessibility". The thermodynamic entropy is defined as (amongst other things) an additive function that is non-decreasing during these transitions. They require that the initial and final states of the system be in equilibrium, in the sense that the system has relaxed to a stationary state but they make it very clear that the process in between need not be slow nor gentle, in fact it can arbitrarily violent (e.g. see Lieb \& Yngvason's ``adiabatic gorilla"). Hence there is in principle no restriction on the coupling strengths under these asymptotic conditions.

One might however worry about internal coupling strengths in the heat flow model. There, weak coupling would imply that the final states after the unitary interaction are still approximately locally diagonal in $H_A, H_B$ whereas strong coupling corresponds to final states that are not diagonal in $H_A, H_B$. We do not assume, a priori, any particular coupling strength; the governing equation \eqref{QTI} is independent of coupling strength. The systems A and B can potentially go to thermal or non-thermal states after the interaction because, as mentioned above, the free energy functional that governs the directionality of the process is defined over all state space (not just Gibbs states), what matters is only that the initial states are locally thermal.

In this sense such a coupling concern is in allowing the final reduced states to not be Gibbsian, leading to $[\rho^f_A, H_A] \neq 0$ and similarly for subsystem B. However this is not necessarily a problem when we remember that the total system AB is isolated, i.e. after the interaction, there is no thermal bath for it to equilibrate with and once the particles are separated after scattering, the reduced states have fixed energies with perhaps coherent oscillations in their energy eigenbasis (even though it might be the case that we cannot ascribe any notion of local temperature - in fact we don't need to if interested solely in questions of energy exchange). What is different in the strong coupling case is that the von Neumann entropy does not equal the thermodynamic entropy any longer (if one defines the thermodynamic entropy to be for the state having energy $\Tr[\rho H]$ but which maximises von Neumann entropy, defined thus in \cite{QMaxWork}). Despite this, in the event of such non-static outcomes, familiar entropic and thermodynamic behaviour is still observed with respect to the initial equilibrium states. This behaviour stems once again from the directionality implied in Eq. \eqref{QTI}, itself born from the positivity of the relative entropy with respect to the initial Gibbs states.

\subsection{\label{secCollision}Towards a Generalised Collision Model}

In an attempt to explain equilibration between two systems, Partovi \cite{PartoviOld} devised a collision model between two particles. We outlined its principles in example 3 in Ref. \cite{FirstPaper} and claimed that when the two systems are qubits, one of Partovi's restrictions can be weakened. Here we elaborate on why this is true and furthermore we explain why in the higher dimensional case, Partovi's initial assumptions must hold.

The fundamentals of the collision process is summarised here: initially the two systems are described by a joint product state $\rho$, they interact via a global energy conserving unitary and then they decorrelate, forgetting any correlations that are produced and rendering the process irreversible (this moves the system to a new unitary orbit). More compactly:
$\rho = \rho_A \otimes \rho_B \rightarrow \rho' = U \rho U^{\dag} \rightarrow \rho'_A \otimes \rho'_B$.
At the end of these steps, the local entropies and global entropy has increased, $\Delta S_A + \Delta S_B \geq 0$, $\Delta S \geq 0$. This whole procedure is then reiterated until the entropies are maximised and the systems reach equilibrium.

We propose a modified version of the model where, instead of the systems completely decorrelating after the unitary interaction, the joint state of the system dephases to a classically correlated state by losing its ``off-diagonal'' components (again this process transfers the system to a new unitary orbit). Note that this requires the introduction of local bases. This process may be more physically meaningful because dephasing occurs faster than decorrelation in real systems.

Let $\sigma, \sigma'$ be two minimum QMI states on some unitary orbits. Then does the process $\sigma \rightarrow \rho' = U \sigma U^{\dag} \rightarrow \sigma'$ achieve the same entropic increase as the original Partovi scheme?
$\sigma'$ is the dephased $\rho'$.

\subsubsection{Two qubits, $d_A = d_B = 2$}

The total Hamiltonian for the system can be taken as $H =  H_A + H_B = \ket{1}\bra{1}_A \otimes \mathbb{I}_B + \mathbb{I}_A \otimes \ket{1}\bra{1}_B = \ket{01}\bra{01} + \ket{10}\bra{10} + 2\ket{11}\bra{11}$. It has an energy $E = 1$ degenerate subspace spanned by $\{\ket{01},\ket{10}\}$. The most general form of the unitary that commutes with this Hamiltonian (in a loose notation) is $U_H = \Pi_{E=0} + U_{E=1} + \Pi_{E=2}$, where $\Pi_{E=0} = \ket{00}\bra{00}, \Pi_{E=2} = \ket{11}\bra{11}$ and $U_{E=1}$ is a general $2\times 2$ unitary rotating in the energy degenerate subspace. Without loss of generality (in terms of action on the marginal eigenvalues), $U_H$ need only rotate in a two-dimensional plane of the effective $\{\ket{01},\ket{10}\}$ Bloch sphere. Thus $U_H$ is given by $U^{(o)}$ in equation (\ref{Uodd}).

The initial minimally correlated state is $\sigma = \lambda_1 \ket{00}\bra{00} + \lambda_2 \ket{01}\bra{01} +\lambda_3 \ket{10}\bra{10} +\lambda_4 \ket{11}\bra{11}$. It is located at the minimally correlated vertex of its reduced state spectra polygon $\mathcal{R}(\sigma)$, c.f. where $\rho_{min}$ is located in figure \ref{figLALBfgh} - at $R_1$ or $R_2$. Its energy conserving region $\mathcal{R}_E(\sigma)$ therefore is just the line $\lambda_A + \lambda_B = E = \lambda_2 + \lambda_3 + 2\lambda_4$, i.e. the lines $f_1$ or $f_2$ in figure \ref{figLALBfgh}, and so $U_H$ (a SECU) evolves $\sigma$ to all states in $\mathcal{R}_E(\sigma)$. The fact that $[ U_H, H] = 0$ means not only that energy is conserved, but also that the reduced states will remain diagonal in their energy eigenbases and, since they are qubits, they will remain thermally distributed (it is always possible to identify the qubit eigenvalues with a Gibbs distribution with respect to $H_A$ and $H_B$) and so one can always define local temperatures.

The first step in the collision process is $\sigma \rightarrow \rho' = U_H \sigma U_H^{\dag}$ and the final state $\rho'$ ends up somewhere along the constant energy line forming $\mathcal{R}_E(\sigma)$ and looks like
\begin{widetext}
\begin{align}
\label{eqrhominSECU}
\rho' =
\frac{1}{2} \left(
  \begin{array}{cccc}
    \lambda_1 & 0 & 0 &0  \\
    0 & p\lambda_2 + (1-p)\lambda_3 & -\sqrt{p(1-p)}(\lambda_2 - \lambda_3) & 0 \\
    0 & -\sqrt{p(1-p)}(\lambda_2 - \lambda_3) & (1-p)\lambda_2 + p\lambda_3 & 0 \\
    0 & 0 & 0 & \lambda_4 \\
  \end{array}
\right)
\end{align}
\end{widetext}
with $p = \cos^2{\theta}$. Now decoherence occurs in the computational basis on the global state and the final state is
\begin{align}
\label{eqdephase}
\sigma' =
\frac{1}{2} \left(
  \begin{array}{cccc}
    \lambda_1 & 0 & 0 &0  \\
    0 & p\lambda_2 + (1-p)\lambda_3 & 0 & 0 \\
    0 & 0 & (1-p)\lambda_2 + p\lambda_3 & 0 \\
    0 & 0 & 0 & \lambda_4 \\
  \end{array}
\right)
\end{align}
The dephasing process moves the state of the system off the original unitary orbit because $S(\sigma') \geq S(\rho') $.
To show $\sigma'$ is still a minimally correlated state in its orbit we must check the ordering of its eigenvalues. It is readily seen that if
$\lambda_1 \geq \lambda_2 \geq \lambda_3 \geq \lambda_4$ then, since $0 \leq p \leq 1$, it is true that $\lambda_1 \geq p\lambda_2 + (1-p)\lambda_3 \geq (1-p)\lambda_2 + p\lambda_3 \geq \lambda_4$ or $\lambda_1 \geq (1-p)\lambda_2 + p\lambda_3 \geq p\lambda_2 + (1-p)\lambda_3 \geq \lambda_4$ and both of these orderings are valid minimally correlated states (one ordering is obtained from the other by a swap of the states of A and B which is a symmetry of the QMI).

The combined transformation on the state can be  written as a stochastic map $\mathcal{M}:\sigma \rightarrow \sum_{k=1}^4 \Pi_k U_H \sigma U_H^{\dag} \Pi_k$, where $\Pi_k$ is a projector onto the joint basis $\ket{i_Aj_B}$, $k=(i_A,j_B)$ and $i_A,j_B=0,1$. After one collision, the map on the initial classically correlated state $\sigma$ transforms it to the state $\sigma'$ in equation (\ref{eqdephase}). After $n$ collisions, the resulting state is
\begin{align}
\label{eqdephase}
\mathcal{M}^n(\sigma) = \sigma^{(n)} =
\frac{1}{2} \left(
  \begin{array}{cccc}
    \lambda_1 & 0 & 0 &0  \\
    0 & \lambda^{(n)}_2 & 0 & 0 \\
    0 & 0 &\lambda^{(n)}_3 & 0 \\
    0 & 0 & 0 & \lambda_4 \\
  \end{array}
\right)
\end{align}
where $$\lambda^{(n)}_2 = p^{(n-1)}\lambda^{(n-1)}_2 + (1-p^{(n-1)})\lambda^{(n-1)}_3$$ $$\lambda^{(n)}_3 = (1-p^{(n-1)})\lambda^{(n-1)}_2 + p^{(n-1)}\lambda^{(n-1)}_3$$

It is simple to see that the intervals satisfy $$\left[\lambda^{(n)}_2,\lambda^{(n)}_3\right]  \in \left[\lambda^{(n-1)}_1,\lambda^{(n-2)}_2\right] \in \dots \in \left[\lambda_1,\lambda_2\right] $$ hence after each collision, the difference between the middle two eigenvalues $\lambda^{(n)}_2, \lambda^{(n)}_3$ decreases and $\lim_{n\rightarrow \infty} |\lambda^{(n)}_2-\lambda^{(n)}_3| = 0$. Also the ordering of the eigenvalues on every new orbit (after dephasing) preserve the minimally correlated form. Thus after very many collisions, the local and global entropies are maximised, the system's reduced states have the same spectra and so equilibrate to the same temperature. This means that it is not necessary for qubits to forget the correlations generated by the unitary interaction; if they only dephase to a classically correlates state, they will still equilibrate to the same temperature.

\subsubsection{The qubit-qutrit case: $d_A = 2, d_B = 3$}

This property does not unconditionally carry over to higher dimensions, which we now demonstrate by looking at the $d_A = 2, d_B = 3$ case.

The Hamiltonian of the total system now is $H  = \ket{01}\bra{01} + \ket{10}\bra{10} + 2(\ket{11}\bra{11} + \ket{02}\bra{02}) + 3\ket{12}\bra{12}$. Now there exists an energy $E = 1$ degenerate subspace spanned by $\{\ket{01},\ket{10} \}$ and an $E=2$ subspace spanned by $\{\ket{11},\ket{02} \}$. Accordingly, the most general form of the unitary which commutes with $H$ is $U_H = \Pi_{E=0} + U_{E=1} + U_{E=2} + \Pi_{E=3}$ where $\Pi_{E=0} = \ket{00}\bra{00}, \Pi_{E=3} = \ket{12}\bra{12}$ and $U_{E=1}, U_{E=2}$ are $2 \times 2$ unitaries rotating independently in their own energy degenerate subspaces.
However unlike the two qubit case, these energy conserving rotations are not ``symmetries'' of the QMI, in the sense that after the unitary operation and the dephasing, the system does not return to a minimally correlated state. For example, let the initial state be $\sigma = \lambda_1 \ket{00}\bra{00} + \lambda_2 \ket{01}\bra{01} + \lambda_3 \ket{02}\bra{02} +\lambda_4 \ket{10}\bra{10} + \lambda_5 \ket{11}\bra{11} + \lambda_6 \ket{12}\bra{12}$, with $\lambda_1 \geq \lambda_2 \geq \lambda_3 \geq \lambda_4 \geq \lambda_5 \geq \lambda_6$; we pick a spectrum $ \Lambda = \{\lambda_i\}_{i=1}^6$ such that this is the its minimally correlated form, see section \ref{gendim}.
A permitted energy conserving transformation in the first step of the collision process is one whose only action is to swap $\lambda_2$ and $\lambda_4$ in $\sigma$.
After dephasing (which in this case has no effect) the state is now $\rho'$ which is equal to $\rho$ up to a permutation of the eigenvalues on the diagonal. This particular permutation raises the QMI of the state $I(\rho') \geq I(\sigma)$ (see section \ref{secPutintoYT}), hence $\rho'$ is no longer a minimally correlated state on its orbit (which here is the same orbit as $\sigma$). This means that in the next iteration of the collision process one can find an energy conserving unitary that decreases $S_A, S_B$ (for example, just the inverse of the unitary that was initially applied to $\sigma$) which goes against the assumptions of Partovi.

To summarise, if we are to concur that systems tend to equilibrate via the Partovi collision model, where at every step the local entropies are non-decreasing, then (apart from two qubit systems) the two colliding systems must decorrelate to a product state after every interaction. In systems with $d > 4$, swaps of energy level populations that preserve energy do not preserve correlations. A corollary of this is given a Hamiltonian $H = H_A + H_B$ and an energy $E = \Tr(\rho H)$, the set of classically correlated states $\{ \rho \}$ that are diagonal in $H$ with energy $E$ have a range of QMI values $\Delta I_{E,H}$. This is a special case of the energy conserving condition analysed in section \ref{secEcons} above, but restricted to classical states.

\subsection{\label{secCharlieDemon}Global demons and local observers}

\begin{figure}[h!]
\includegraphics[width=3in]{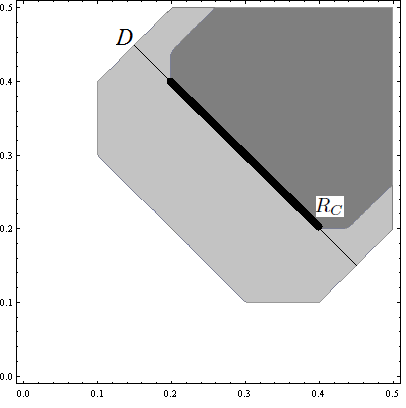}
\caption{The darker (embedded) grey region is the set Charlie thinks he has unitary access to $\mathcal{R}(\Lambda^{(C)})$. His description of the state is located and the minimally correlated vertex $R_C$ of this set, the energy conserving region $\mathcal{R}_E(\Lambda^{(C)})$ for Charlie's state is the thick black line. The entire (light plus dark) grey region is the set of states the demon has access to, $\mathcal{R}(\Lambda^{(D)})$ (a different unitary orbit), and its description of the system's marginals coincides with Charlie's, but the state the demon detects $\rho^{(D)}$ is now some correlated state in $\mathcal{R}(\Lambda^{(D)})$. The demon's equal energy region is on and below the thick and thin black line. To confuse Charlie it evolves the state $\rho^{(D)}$ to $\gamma^{(D)}$ which is situated at D.}
\label{figCharlieDemon}
\end{figure}

Our analysis of the correlations in a bipartite state can be phrased in terms of an paradoxical scenario in which a global Maxwell demon can confuse an observer Charlie who can only measure local observables.
Charlie is handed pairs of qubits he knows are described by states $\rho_A = (1-\lambda_A)\ket{0}\bra{0} + \lambda_A\ket{1}\bra{1}, \rho_B=(1-\lambda_B)\ket{0}\bra{0} + \lambda_B\ket{1}\bra{1}$, $0\leq\lambda_A,\lambda_B\leq\frac{1}{2}$ with Hamiltonians $H_A = \ket{1}\bra{1}, H_B=\ket{1}\bra{1}$, and he can makes measurements in these energy bases to find the expectation values $\langle H_A \rangle, \langle H_B \rangle$. Charlie knows the states of the qubits $\rho_A,\rho_B$ and therefore also the initial energies $E_A=\Tr[\rho_A H_A] = \lambda_A, E_B=\Tr[\rho_B H_B]=\lambda_B$. What Charlie does not know is the global state $\rho$ of the two systems A and B. Without any knowledge of correlations in the system, his description of it is local, that is, Charlie's most unbiased estimate of the system state, given his knowledge, is $\rho^{(C)} = \rho_A \otimes \rho_B$, with QMI $I(\rho^{(C)})=0$. He decides to interact A and B via an energy conserving unitary interaction, as in the earlier heat flow model. If any correlations are present, he may see negative heat flow as an indicator of them. His description of the system $\rho^C$ is the minimally correlated state in its unitary orbit with spectrum $\Lambda^{(C)} = \{(1-\lambda_A)(1-\lambda_B), (1-\lambda_A)\lambda_B,\lambda_A(1-\lambda_B),\lambda_A\lambda_B\}$, therefore it must be situated at the minimally correlated vertex $R_C$ in its reduced state spectrum set $\mathcal{R}(\Lambda^{(C)})$, see figure \ref{figCharlieDemon}. The equal energy region $\mathcal{R}_E(\Lambda^{(C)})$ for this state is the line that it is situated on, highlighted in the figure. All the states corresponding to $\mathcal{R}_E(\Lambda^{(C)})$ are obtainable by performing a strong energy conserving unitary (SECU) transformation on $\rho^{(C)}$, such as $U^{(o)}(\theta)$ given in equation (\ref{Uodd}). Charlie therefore interacts the qubits so that their evolution is governed by $U^{(o)}(\theta)$ and finds that he \textit{always} observes normal heat flow. He concludes that his description of the system is correct.

Now we look from the demon's perspective: it has access to non-local observables and so decribes the system via a correlated state $\rho^{(D)}$ with the property $\Tr_A(\rho^{(D)}) = \Tr_A(\rho^{(C)}) = \rho_B$ and $\Tr_B(\rho^{(D)}) = \Tr_B(\rho^{(C)}) = \rho_A$. To make this example more concrete, we specify the state $\rho^{(D)}$ with spectrum $\Lambda^{(D)} = \{\lambda_1, \lambda_2,\lambda_3,\lambda_4\}$, and $\lambda_1\geq \lambda_2\geq\lambda_3\geq\lambda_4$, to be
\begin{widetext}
\begin{align}
\rho^{(D)} =
\left(
  \begin{array}{cccc}
    \frac{1}{2}(\lambda_1 + \lambda_4 + (\lambda_1 - \lambda_4)\cos\phi) & 0 & 0 & \frac{1}{2}(\lambda_4 - \lambda_1)\sin\phi \\
    0 & \lambda_2 & 0 & 0 \\
    0 & 0 & \lambda_3 & 0 \\
    \frac{1}{2}(\lambda_4 - \lambda_1)\sin\phi & 0 & 0 & \frac{1}{2}(\lambda_1 + \lambda_4 - (\lambda_1 - \lambda_4)\cos\phi) \\
  \end{array}
\right)
\end{align}
\end{widetext}
and for the demon's reduced states to be compatible with Charlie's we demand that $\frac{1}{2}(1-\lambda_2 + \lambda_3 - (\lambda_1 - \lambda_4)\cos\phi) = \lambda_A$ and $\frac{1}{2}(1+\lambda_2 - \lambda_3 - (\lambda_1 - \lambda_4)\cos\phi) = \lambda_B$. These equalities are satisfied, for instance, when $\Lambda^{(D)} = \{0.6,0.3,0.1,0\}$ in which case $\lambda_A = \lambda_B - 0.2$ and $\cos\phi = \frac{0.4-\lambda_A}{0.3}$, last condition requires $\lambda_A \geq 0.1$. Picking e.g. $\lambda_B = 0.4$ gives valid $\lambda_A = 0.2, \cos\phi = \frac{2}{3}$, and for this choice of values $I(\rho^{(D)}) \approx 0.4$, so the demon's state is correlated, but has the same marginals as Charlie. However under the strong energy conserving unitary that Charlie is applying, $\rho^{(D)}$ evolves to the state $\rho(\lambda_1,\lambda_2,\lambda_3,\cos\theta,\frac{2}{3})$, see equation (\ref{rhothetaphi}), using results from section $\ref{secAnalysisR}$, one can see that the QMI is concave over the domain of $\theta$ and its minima occurs at the extrema, i.e. when $\theta = 0 (\pi)$, which is $\rho^{(D)}$. So $I(U^{(o)}(\theta)\rho^{(D)}U^{(o)}(\theta)^{\dag})\geq I(\rho^{(D)}) \, \forall \theta$ and the heat flow model dictates that normal heat flow will always occur for this special correlated state $\rho^{(D)}$ under the action of $U^{(o)}$, this is why Charlie always sees normal heat flow, even though the system is in reality correlated: his unitary $U^{(o)}$ does not ``reveal'' the correlations.

However, since $\rho^{(C)}$ and $\rho^{(D)}$ have different spectra, they are located on different unitary orbits and define different marginal eigenvalue sets $\mathcal{R}(\Lambda^{(C)})$ and $\mathcal{R}(\Lambda^{(D)})$. Sticking with the values chosen for all the eigenvalues above, we see (pictorially in figure \ref{figCharlieDemon}) that $\mathcal{R}(\Lambda^{(C)}) \subset \mathcal{R}(\Lambda^{(D)})$. Furthermore, the equal energy region from the demon's point of view is everything one and below the thick and thin black line, i.e. $\mathcal{R}_E(\Lambda^{(C)}) \subset \mathcal{R}_E(\Lambda^{(D)})$. It is this fact that gives the demon his advantage: because of its knowledge of the correlations and the larger constant energy region, it can now transform the system to a state in $\mathcal{R}_E(\Lambda^{(D)})$ that is less correlated and transfers heat from the cold to the hot qubit. This cannot be done using a SECU, instead the demon must use a weak energy conserving unitary (WECU) that preserves total energy energy on average but does not commute with the total Hamiltonian $H = H_A + H_B$. The demon decides to interfer with Charlie's experiment and, while Charlie is not looking, switches on a demonic unitary interaction, producing a state
\begin{widetext}
\begin{align}
\gamma^{(D)} =
\left(
  \begin{array}{cccc}
    \frac{1}{2}(\lambda_1 + \lambda_3 + (\lambda_1 - \lambda_3)\cos\phi') & 0 & 0 & \frac{1}{2}(\lambda_4 - \lambda_3)\sin\phi' \\
    0 & \lambda_2 & 0 & 0 \\
    0 & 0 & \lambda_4 & 0 \\
    \frac{1}{2}(\lambda_3 - \lambda_1)\sin\phi' & 0 & 0 & \frac{1}{2}(\lambda_1 + \lambda_3 - (\lambda_1 - \lambda_3)\cos\phi') \\
  \end{array}
\right)
\end{align}
\end{widetext}
By taking partial traces, one finds that this state has thermal marginals (not necessary for our heat flow model but included to demonstrate the importance of WECUs), and $\mathrm{spec}(\gamma^{(D)}) = \Lambda^{(D)}$ hence it is on the unitary orbit of $\rho^{(D)}$. The total energy of $\gamma^{(D)}$ is $\Tr[(H_A + H_B)\gamma^{(D)}] = 1 + (\lambda_3 - \lambda_1)\cos\phi'$, to satisfy energy conservation $\Tr[(H_A + H_B)\gamma^{(D)}] = \Tr[(H_A + H_B)\rho^{(D)}] = 0.6$, this sets $\cos\phi' = \frac{4}{5}$. The final QMI is $I(\gamma^{(D)}) \approx 0.3 < I(\rho^{(D)})$  and system A, which started with energy 0.2, now has energy 0.15 and system B, which started with energy 0.4, now has energy 0.45. Hence the cold system has got colder, the hot system has got hotter and energy has remained constant \textit{and} we have chosen an interaction that results in local thermal states so that we can talk about temperature (hotness and coldness).

When Charlie makes his usual energy measurements of the state that the demon has tampered with, he observes anomalous heat flow and is very confused because he thought he had correctly described the system as being uncorrelated so only expected normal heat exchange (all of his experiments led him to believe this). The observation of heat flow from cold to hot with no other change in the system (constant energy and constant correlations) looks like a real paradox and potentially a violation of the second law. This shows that thermodynamics is only properly defined with respect to the particular class of observables that are used to prepare and manipulate the system. In turn, thermodynamic entropy must be defined with respect to this class if one wants to universally uphold the second law.

\section{\label{gendim}Determining $\rho_{min}$ in the general dimensional case}

We now return to the problem of determining the absolute minimum attainable correlation on a given unitary orbit for a bipartite system of arbitrary size.
For two qubits we found $\rho_{min}$ by directly minimising $I(\rho)$ over the unitary orbit. In fact, it is not necessary to solve the whole quantum marginal problem in order to find $\rho_{min}$ - sufficient is the solution to the simpler problem of finding the convex hull of the unitary orbit.

Even though the domain is now extended to the convex hull of $\mathcal{O}_\rho$, we initially ignore the total entropy $S(\rho)$ in our optimisation and focus on just the variation of the function $G(\sigma) := S(\sigma_A) + S(\sigma_B)$ over all states $\sigma$ in the convex hull. Omitting $S(\rho)$ is not a problem because as it turns out the minimum of $G(\sigma)$ over the convex hull will be given by a state $\rho_{min}$ on the unitary orbit of $\rho$.

A state is in the convex hull $\mathcal{C}_\rho$ of $\mathcal{O}_\rho$ if it can be written in the form $\sigma = \sum_i p_i U_i \rho U_i^\dag$ for a set of unitaries $\{U_i \}$ and with $\sum_i p_i = 1, p_i > 0 \, \forall \, i$. An equivalent condition is \cite{Wehrl} $\mathrm{spec}(\sigma) =: \mathbf{\Sigma} \prec \mathrm{spec}(\rho) = \mathbf{\Lambda}$ where $\mathbf{\Sigma},\mathbf{\Lambda}$ are vectors of the eigenvalues of $\sigma,\rho$ respectively and we have introduced the majorization relation as a partial order on vectors. A vector $\mathbf{p}$ is said to majorise another vector $\mathbf{q}$, denoted $\mathbf{q} \prec \mathbf{p}$, for each $k = 1, \dots, d$ the following holds:
\begin{align}
\sum^k_{i=1} q^{\downarrow}_i \leq \sum^k_{j=1} p^{\downarrow}_j.
\end{align}
$d$ is the dimension of $\mathbf{p,q}$, $p_i, q_i \geq 0$ and equality holds only when $k=d$ \cite{MarshOlkin}.
In Ref. \cite{Bravyi} it is shown that the eigenvalues of all marginal states in $\mathcal{C}_\rho$ are given by marginals of probability distributions $\mathbf{P}$ majorised by $\mathbf{\Lambda}$. This reduces the minimisation down to a classical problem: the function $G(\sigma) = S(\sigma_A) + S(\sigma_B) = H(P_A) +  H(P_B) =: G(P)$, where $P_A,P_B$ are marginals of $P$ and are sets containing elements of the vector $\mathbf{P}$, is now to be minimised over the convex set of probability vectors satisfying $\mathbf{P} \prec \mathbf{\Lambda}$. It is straight forward to show that $G(P)$ is concave on this set of vectors.
Therefore minima occur at the extrema of this set, that is when $\mathbf{P} = \Pi({\mathbf{\Lambda}})$ \cite{Bravyi}, where $\Pi$ is a permutation on the components of $\mathbf{\Lambda}$. The states that these extrema correspond to are minimally correlated states $\rho_{min} \in \mathcal{C}_\rho$ fulfilling the property $\mathrm{diag}(\rho_{min}) = \Pi(\Lambda)$, thus they are in fact on the unitary orbit: $\rho_{min} \in \mathcal{O}_\rho$.

So we have reduced the problem down to the following: the minimum of the QMI on a unitary orbit has a value of $I_{min} = H([\Pi(\Lambda)]_A) + H([\Pi(\Lambda)]_B) - H(\Lambda)$, and the state that corresponds to this is a classically correlated one
$$
\rho_{min} = \sum_{j,k=1}^{d_A,d_B} \lambda_{j,k} \ket{j}_A\bra{j}\otimes\ket{k}_B\bra{k}
$$
where $\{\ket{j}_A\}, \{\ket{k}_B\}$ are local bases and $\lambda_{j,k}$ is a reindexing of the eigenvalues $\lambda_i$: $\lambda_{j,k} = \lambda_{i=(j-1)d_B + k}$.

As in the two qubit case, the ordering of the eigenvalues matters, we now turn our attention to finding the permutation $\Pi$ that gives the minimum QMI.

\subsection{\label{2qubitConvHull}The convex hull $\mathcal{C}_\rho$ of the two qubit system}

\begin{figure}[h!]
\includegraphics[width=3.5in]{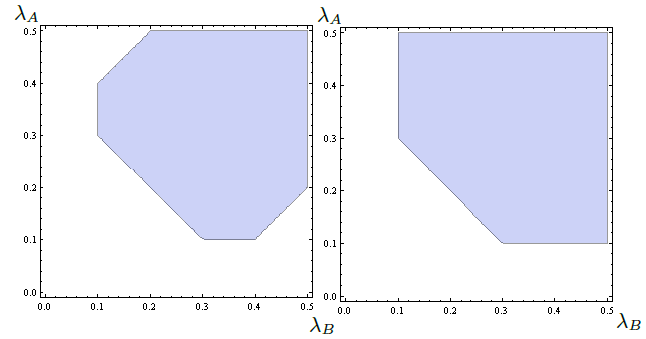}
\caption{Comparing the set of reduced state spectra $\mathcal{R}$, on the left, and its convex hull, on the right. Although $\mathcal{R}$ on its own looks like a convex polygon, one must remember that in fact the eigenvalues $\lambda_A, \lambda_B$ are defined over the range $[0,1]$, whereas here only a portion of this, $[0,\frac{1}{2}]$, is represented. If we imagine (using symmetry) the $\mathcal{R}$ over the full range $[0,1]$, which is the domain on which the entropies $H(\lambda_A), H(\lambda_B)$ are defined, we see that it is not convex.}
\label{figConvHullLALB}
\end{figure}

Before we move on it is constructive to regard what the convex hull looks like for the two qubit system. In section \ref{rhomin} we found the minimum of the QMI ``manually'' because the domain $\mathcal{R}$ was not convex. The convex hull of a typical $\mathcal{R}$ is shown next to $\mathcal{R}$ in figure \ref{figConvHullLALB} for $0 \leq \lambda_A, \lambda_B \leq \frac{1}{2}$. It is evident now what reduced states are absent in the unitary orbit: unitary evolution constrains the reduced states to be closer in mixedness than the map that is a convex combination of unitaries. The constancy of the joint state spectrum acts as a ``tension'' between the reduced states, restricting the marginals to being of similar Bloch vector length. In fact, the inequality (Eq. \ref{ineq4}) that cuts away the part of the convex hull so as to give $\mathcal{R}$ is clearly the defining feature for the marginals of $\mathcal{O}_\rho$ and is the hardest to find. As Bravyi puts it, this task is ``rather formidable'' \cite{Bravyi}. It is this lack of convexity that contributes to why the quantum marginal problem is so difficult.

\subsection{\label{secMinOverPerms}Minimising the $H(\Lambda_A) + H(\Lambda_B)$ over permutations of the elements of $\Lambda$}

From now on, unless explicitly stated, we ignore the joint entropy $H(\Lambda)$ as we will be looking at individual orbits.
Let the components of the vector of eigenvalues $\mathbf{\Lambda}$, which is essentially a bipartite probability distribution, be
arranged in non-increasing order, $\lambda _{i}\geq \lambda _{i+1} \, \forall \,
i=1,\dots ,d-1$.
The $d = d_A d_B$ eigenvalues can be arranged in a table $\mathbb{T}$,
such that the first $d_{B}$ elements of $\mathbf{\Lambda}$ occupy the first row of $\mathbb{T}$, the
second $d_{B}$ elements reside in the second row, and so on until we occupy $d_{A}$
rows and $d_{B}$ columns. This setup makes it easier to see how permutations of
these elements affect the marginal probability vectors $\mathbf{\Lambda}_A, \mathbf{\Lambda }_{B}.$

Each eigenvalue $\lambda _{i}$ is identified with
a table entry $\tau _{r,c}$, the rows are
labelled by $r=1,\dots ,d_{A}$ and columns by $c=1,\dots ,d_{B},$ and we
denote by
$\mathbf{R}^{(r)}$ $\left( \mathbf{C}^{(c)}\right) $ the vector of table
components in the $r$-th row$\ (c$-th column). The table below demonstrates this: it shows just one way of arranging the $\lambda_i$s,
\begin{eqnarray*}
&&%
\begin{tabular}{l||l|l|l|l}
$r$ $\backslash$ $c$ & 1 & 2 & $\dots $ & $d_{B}$ \\
\hline \hline
1 & $\lambda _{1}$ & $\lambda _{2}$ & $\dots $ & $\lambda _{d_{B}}$ \\
\hline
2 & $\lambda _{d_{B}+1}$ & $\lambda _{d_{B}+2}$ & $\dots $ & $\lambda
_{2d_{B}}$ \\
\hline
$\vdots $ & $\vdots $ & $\vdots $ & $\ddots $ & $\vdots $ \\
\hline
$d_{A}$ & $\lambda _{d_{B}\left( d_{A}-1\right) +1}$ & $\lambda
_{d_{B}\left( d_{A}-1\right) +2}$ & $\dots $ & $\lambda _{d_{A}d_{B}}$%
\end{tabular}
\\
&=&
\begin{tabular}{l||l|l|l|l}
$r$ $\backslash$ $c$ & 1 & 2 & $\dots $ & $d_{B}$ \\
\hline \hline
1 & $\tau _{1,1}$ & $\tau _{1,2}$ & $\dots $ & $\tau _{1,d_{B}}$ \\
\hline
2 & $\tau _{2,1}$ & $\tau _{2,2}$ & $\dots $ & $\tau _{2,d_{B}}$ \\
\hline
$\vdots $ & $\vdots $ & $\vdots $ & $\ddots $ & $\vdots $ \\
\hline
$d_{A}$ & $\tau _{d_{A},1}$ & $\tau _{d_{A},2}$ & $\dots $ & $\tau
_{d_{A},d_{B}}$%
\end{tabular}%
= \mathbb{T}
\end{eqnarray*}%
and%
\begin{eqnarray*}
\mathbf{R}^{(r)} 
=\left( \tau _{r,1,}\dots ,\tau _{r,d_{B}}\right) \\
\mathbf{C}^{(c)} 
=\left( \tau _{1,c,}\dots ,\tau _{d_{A},c}\right)
\end{eqnarray*}%

\bigskip When written in tabular form, it is easy to calculate the marginal probabilities vectors (MPVs) of $\mathbf{\Lambda}$. The MPVs for systems A and B are denoted $\mathbf{a}$  and $ \mathbf{b}$ respectively, their components
are just the sums of the elements in the row and column vectors of the table $\mathbb{T}$
\begin{eqnarray*}
\mathbf{a} &=&\left( \sum_{i=1}^{d_{B}} R^{(1)}_{i},\dots
,\sum_{i=1}^{d_{B}}R^{(r)}_{i},\dots ,\sum_{i=1}^{d_{B}}
R^{(d_{A})}_{i}\right) \\
\mathbf{b} &=&\left( \sum_{i=1}^{d_{A}}C^{(1)}_{i},\dots
,\sum_{i=1}^{d_{A}}C^{(s)}_{i},\dots ,\sum_{i=1}^{d_{B}}
C^{(d_{B})}_{i}\right)
\end{eqnarray*}%
$\mathbf{a}$ and $\mathbf{b}$ are equivalent to what we previously called $%
\mathbf{\Lambda }_{A}$ and $\mathbf{\Lambda }_{B}.$ We use $\Pi(\mathbb{T})$ to mean the permutation on the table elements that corresponds to $\Pi(\Lambda)$. A reordering $\Pi$ changes the MPVs
$\mathbf{a}$ and $\mathbf{b}$ and hence alters $H\left( \mathbf{a}\right)
+H\left( \mathbf{b}\right)$; we seek the permutation $\Pi$  that gives the minimum of the sum of entropies.

\subsubsection{\label{secPutintoYT}Sorting rows and columns into decreasing order}

We define two sorting algorithms on the tables: $\mathfrak{A}^{row}$ ($\mathfrak{A}^{col}$) sorts rows (columns) into descending order.
Consider all $d!$ tables of $d$ elements, which are denoted $\mathbb{T}_d$, applying these operations in tandem to all tables generates a subset $Y_d$ of tables with all their elements arranged in decreasing order in their rows and columns, $Y_d := \{ \mathbb{T}^{\downarrow} = \mathfrak{A}\mathbb{T}_d : \forall \, \mathbb{T}_d, \mathfrak{A}:=\mathfrak{A}^{row}\mathfrak{A}^{col} \} = \{ \mathbb{T}^{*\downarrow} = \mathfrak{A}^*\mathbb{T}_d : \forall \, \mathbb{T}_d, \mathfrak{A}^*:=\mathfrak{A}^{col}\mathfrak{A}^{row}  \}$.
Note however that in the case when $d_A = d_B$, we define this set as \textit{not} containing tables that are transposes of another table in $Y_d$ even though these are generated by the sorting operations, that is $\mathbb{T}^{\downarrow} \in Y_{(d_A)^2}$ and $(\mathbb{T}^{\downarrow})^T \not\in Y_{(d_A)^2}$. This is because they have the same QMI (a transposition on a table is a symmetry of the QMI, see section \ref{secTableSymmQMI}).

The table that dictates how the eigenvalues should be arranged in order to obtain the minimum mutual information is in the set $Y_d$.
To see this, consider an unordered table $\mathbb{T}_d$. The entries of row MPVs $\mathbf{a}$ and $\mathbf{a}^{\downarrow }$ are%
\begin{eqnarray*}
a_{r} &=&\sum_{i=1}^{d_{B}}R^{(r)}_{i} \\
a^{\downarrow } _{r} &:&=\sum_{i=1}^{d_{B}}
R^{(r)\downarrow }_{i}
\end{eqnarray*}%
where $\mathbf{R}^{(r)\downarrow }$ is the $r$-th row vector of $\mathfrak{A}^{col}\mathbb{T}_d$. In Ref. \cite{MarshOlkin} it is shown that $\mathbf{a}$
and $\mathbf{a}^{\downarrow }$ obey a majorization relation $\mathbf{a}\prec \mathbf{a}^{\downarrow }$
and since entropy is Schur concave \cite{MarshOlkin}, this means $H\left( \mathbf{a}\right) \geq H\left( \mathbf{a}^{\downarrow }\right)$.

In addition sorting the rows and obtaining table $\mathfrak{A}\mathbb{T}_d \in Y_d$ leads also to $\mathbf{b}\prec \mathbf{b}^{\downarrow }$, and so $H\left( \mathbf{b}\right) \geq H\left( \mathbf{b}^{\downarrow }\right)$
leading to
\begin{equation*}
H\left( \mathbf{a}\right) +H\left( \mathbf{b}\right) \geq H\left( \mathbf{a}%
^{\downarrow }\right) +H\left( \mathbf{b}^{\downarrow }\right).
\end{equation*}%
Hence the operation $\mathfrak{A}$ (or $\mathfrak{A}^*$)
lowers or keeps constant the sum of entropies and the minimum of the mutual information is always given by a table in $Y_d$.
Note that this also holds if $\mathfrak{A}^{row}, \mathfrak{A}^{col}$ are defined as putting row and column elements into ascending order.

\bigskip

The set of tables that result from this sorting are the
$d_{A}\times d_{B}$ Young tableaux.
The number of these tables is given by the hook formula of
Frame-Thrall-Robinson \cite{HookFTR}
\begin{equation}
\label{NumYTFTR}
N(d_A,d_B) = \frac{(xy)!\Pi_{i=1}^{d_B-1}i!}{\Pi_{j=d_A}^{d_A + d_B -1} j!}
\end{equation}%
and this series grows exponentially: $N(2,2) = 2$, $N(2,3) = 5$, $N(3,3) = 42$, $N(4,4) = 24024$.

Our set of interest is $Y_d$, and because it is defined as not containing transposes in the case when $d_A = d_B$, there are in fact only $\frac{1}{2}N(d_A,d_A)$ tables in $Y_{(d_A)^2}$, these are what we call the ``independent Young tableaux''.

\subsection{\label{secTableSymmQMI}An Aside: Symmetry of the Mutual Information in Terms of Tables}

We noted earlier in section \ref{SettingScene} that the quantum mutual information is invariant under particular transformations, they were local unitary transformations and a swap of subsystem states. We can now talk about this in the language of table element permutations. The notation $I(\mathbb{T})$ symbolizes the mutual information of the probability distribution arranged in the order described by table $\mathbb{T}$.
Transformations on the table $\mathbb{T}$ which are symmetries of the QMI are:

1) Transposition (only for tables where $d_A =d_B$) %
\begin{equation*}
I\left( \mathbb{T}^{T}\right) =I\left( \mathbb{T}\right)
\end{equation*}%
This amounts to swapping the distributions of A and B.

2) Permuting rows: writing the table with row vectors as $\mathbb{T}%
\left[ \mathbf{R}^{(1)},\dots ,\mathbf{R}^{(d_{A})}\right] ,$ and a permutation on the
row vector ordering $\Pi_{\mathbf{R}}$%
\begin{align*}
I\left( \mathbb{T}\right) \equiv I\left( \mathbb{T}\left[ \mathbf{R}%
^{(1)},\dots ,\mathbf{R}^{(d_{A})}\right] \right) =\\
I\left( \mathbb{T}\left[ \Pi_{%
\mathbf{R}}\left( \mathbf{R}^{(1}),\dots ,\mathbf{R}^{(d_{A})}\right) \right] \right)
\end{align*}
This is a local operation on the subsystem B.

3) Permuting columns: writing the table with column vectors as $\mathbb{T}\left[ \mathbf{C}^{(1)},\dots ,\mathbf{C}^{(d_{B})}\right] ,$ and a permutation on
the column vector ordering $\Pi_{\mathbf{C}}$%
\begin{align*}
I\left( \mathbb{T}\right) \equiv I\left( \mathbb{T}\left[ \mathbf{C}%
^{(1)},\dots ,\mathbf{C}^{(d_{B})}\right] \right) =\\I\left( \mathbb{T}\left[ \Pi_{%
\mathbf{C}}\left( \mathbf{C}^{(1)},\dots ,\mathbf{C}^{(d_{B})}\right) \right] \right)
\end{align*}
This is a local operation on the subsystem B.

4)\ Swapping all table components $\tau _{rs}$ from, say, descending to
ascending order: a combination of the above transformations.

\bigskip

These transformations can be classed in terms of a group action.
The symmetry group for $I\left( \rho \right) $ is the permutation group $%
S_{D}$ of order $D!$ acting independently on the rows and columns.
In a $d$-dimensional system there are $d!$ permutations/tables. The symmetries reduce the number of tables by a factor of $d_{A}!d_{B}!$, and an additional factor of 2 if $d_{A}=d_{B}$. Note that the Young tableaux are a further subset of this reduced set. This is discussed in more detail in Ref. \cite{Gary}.

\subsection{\label{MinYT2qubit}Special Case: Minimum mutual information
state when $d_{A}=d_{B}=2$}

\bigskip\ For a two qubit system, there is one independent Young table, from Eq. \eqref{NumYTFTR}, $\frac{1}{2}N(2,2) = 1$, it is
\begin{equation*}
\mathbb{T}_{4}^{\downarrow }=%
\begin{tabular}{l||l|l}
$r$ $\backslash$ $c$ & 1 & 2 \\ \hline\hline
1 & $\lambda _{1}$ & $\lambda _{2}$ \\ \hline
2 & $\lambda _{3}$ & $\lambda _{4}$%
\end{tabular}%
\end{equation*}%
This table minimises $H(\mathbf{a}) + H(\mathbf{b})$
for all $\mathbf{\Lambda}$ and corresponds to the state found in equation (\ref{eqrhomin0}), confirming the derivation in section \ref{rhomin}.

In total there are 4! = 24 tables of this dimensionality, but due to
symmetry of the QMI there are only $24/2^3 = 3$ independent tables, i.e. only three distinct values of the QMI. The tables chosen to represent them are the one given above and
\begin{equation*}
\mathbb{T}_{4}^{(1)}=%
\begin{tabular}{l||l|l}
$r$ $\backslash$ $c$ & 1 & 2 \\ \hline\hline
1 & $\lambda _{2}$ & $\lambda _{1}$ \\ \hline
2 & $\lambda _{3}$ & $\lambda _{4}$%
\end{tabular}, \,
\mathbb{T}_{4}^{(2)}=%
\begin{tabular}{l||l|l}
$r$ $\backslash$ $c$ & 1 & 2 \\ \hline\hline
1 & $\lambda _{3}$ & $\lambda _{1}$ \\ \hline
2 & $\lambda _{2}$ & $\lambda _{4}$%
\end{tabular}%
\end{equation*}%
Note these two tables are not in $Y_4$.

We can order the row and column MPVs of these tables with respect to majorization
\begin{align}
\mathbf{a}(\mathbb{T}_{4}^{(2)} ) \prec \mathbf{a}(\mathbb{T}_{4}^{(1)}) = \mathbf{a}(\mathbb{T}_{4}^{\downarrow })\\
\mathbf{b}(\mathbb{T}_{4}^{(2)} ) = \mathbf{b}(\mathbb{T}_{4}^{(1)}) \prec \mathbf{b}(\mathbb{T}_{4}^{\downarrow })
\end{align}
Writing the QMI of table $\mathbb{T}$ as $I(\mathbb{T})$ we have then that
\begin{align}
I(\mathbb{T}_{4}^{(2)})  \geq I(\mathbb{T}_{4}^{(1)}) \geq I(\mathbb{T}_{4}^{\downarrow })
\end{align}
This gives us a full ranking of the mutual information in the full two qubit quantum space. On a unitary orbit, in increasing order
$$I(min,class) = H(\lambda_3 + \lambda_4) + H(\lambda_2 + \lambda_4) - H(\Lambda)$$
$$I(max,class) = H(\lambda_2 + \lambda_4) + H(\lambda_1 + \lambda_4) - H(\Lambda)$$
$$I(max,sep) \leq \log d_A = 1$$
$$I(max,quant) = 2 - H(\Lambda)$$
where the QMIs are over the set of states that are class = classical, sep = separable, quant = arbitrary quantum. The maximum separable value is ascertained from the upper bound of the classical mutual information over the set of all probability distributions with four elements and assumes $2-H(\Lambda) \geq 1$, or equivalently $H(\Lambda) \leq 1$. This full ordering of the mutual information is possible only for two qubits.

\subsection{The correlations structure in higher dimensions: $d_{A}\geq 2,d_{B}\geq 3$}

For systems of larger dimension, the situation becomes more complex because
there are now multiple Young tableaux from which we must determine the one that is the minimiser for
the mutual information (MI) (the problem is now purely classical).

A general solution for $d > 4$ does not yet exist and, unlike the $2 \times 2$ case, it depends on the choice of $\mathbf{\Lambda }$ itself; every member of the Young tableaux set is optimal for some $\Lambda$. For instance, when $d_A = 2, d_B = 3$, the distribution $\Lambda = \frac{1}{21}\{6,5,4,3,2,1 \}$ minimises the MI when the elements are arranged as in table $\mathbb{T}^{(3)}_6$, whereas if $\Lambda = \frac{1}{33}\{10,9,8,3,2,1 \}$, then the optimal table is $\mathbb{T}^{(1)}_6$, these tables are given in section \ref{sec2x3Tables}.

\begin{figure}[h!]
\includegraphics[width=3.5in]{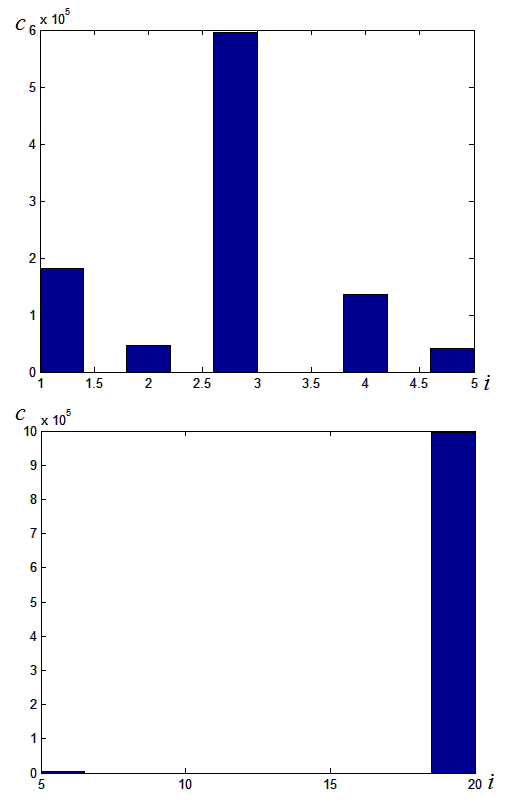}
\caption{Histograms displaying the number of times $c$ a table $\mathbb{T}^{(i)}_{d=d_Ad_B}$ (labelled $i$) minimises the mutual information for $d = 6$ top figure and $d=9$ bottom figure. In each case 1,000,000 full rank probability distributions were generated randomly. Surprisingly the most frequently occurring tables are $\mathbb{T}^{(3)}_{6}$ and $\mathbb{T}^{(19)}_{9}$, in the latter case $\mathbb{T}^{(19)}_{9}$ dominates the histogram.}
\label{figHist3x3YT}
\end{figure}

Below we list explicitly the Young tables for the cases when $d_{A}=2,$ $d_{B}=3$
and $d_{A}=d_{B}=3.$ The set $Y_{d}$ is the collection of ``independent'' (not including transposes)
Young tables $\mathbb{T}_{d}^{i},$ $i=1,\dots \left\vert Y_{d}\right\vert =N $. From now we use the less cumbersome notation $%
i:=\lambda _{i},$ $i=1,\dots ,d.$ Also included are frequency plots in figure \ref{figHist3x3YT} for the tables that minimise the MI for randomly generated probability distributions.

\subsubsection{\label{sec2x3Tables}The distinguished set of tables for $d_{A}=2,$ $d_{B}=3:\ Y_6$}
For the qubit-qutrit case, a direct calculation results in the following set of tables for $Y_6$
\begin{eqnarray*}
\mathbb{T}_{6}^{(1)} =%
\begin{tabular}{l|l|l}
1 & 2 & 3 \\
\hline
4 & 5 & 6%
\end{tabular}, \,
\mathbb{T}_{6}^{(2)} =%
\begin{tabular}{l|l|l}
1 & 2 & 4 \\
\hline
3 & 5 & 6%
\end{tabular}, \,
\mathbb{T}_{6}^{(3)} =%
\begin{tabular}{l|l|l}
1 & 2 & 5 \\
\hline
3 & 4 & 6%
\end{tabular},\\
\mathbb{T}_{6}^{(4)} =%
\begin{tabular}{l|l|l}
1 & 3 & 5 \\
\hline
2 & 4 & 6%
\end{tabular},\,
\mathbb{T}_{6}^{(5)} =%
\begin{tabular}{l|l|l}
1 & 3 & 4 \\
\hline
2 & 5 & 6%
\end{tabular}
\end{eqnarray*}

Curiously, although we cannot deduce which table in $Y_{6}$ gives the minimum MI for a given $\Lambda$, we do know there is a unique table (not in the set $Y_6$) that gives the maximum classical MI. The table is
\begin{eqnarray*}
\mathbb{T}_{6}^{max,class} =%
\begin{tabular}{l|l|l}
1 & 4 & 5 \\
\hline
6 & 3 & 2%
\end{tabular}
\end{eqnarray*}
and this is proved in Ref. \cite{Gary} by introducing a new kind of partial order called ``quasi-majorisation''.

So for $d_A = 2, d_B = 3$, we only partially know the ranking of the mutual information:
$$I(min,class) \le I(max,class) $$
$$I(max,class) = H(\lambda_2 + \lambda_4 + \lambda_6) + H(\lambda_1 + \lambda_6, \lambda_3 + \lambda_4) - H(\Lambda)$$
$$I(max,class) \le I(max,sep) \le \log (\min\{d_A,d_B\}) = 1$$
$$I(max,quant) \leq log 6.$$

Broadly speaking, it is possible for a minimally correlated classical state to be unitarily evolved to the maximally correlated classical state, however this classical state is not necessarily the most correlated state over the set of separable states intersecting the unitary orbit. Thus a further unitary can increase the correlations to form a separable state which has non-zero discord. Beyond this, it may be possible to leave the convex set of separable states and obtain an entangled mixed state.

\subsubsection{\label{sec3x3Tables}The distinguished set of tables for $d_{A}=d_{B}=3:Y_{9}$}

For the case of two qutrits we find that the set $Y_9$ is given by

\begin{eqnarray*}
\mathbb{T}_{9}^{(1)} &=&%
\begin{tabular}{l|l|l}
1 & 2 & 3 \\ \hline
4 & 5 & 6 \\ \hline
7 & 8 & 9%
\end{tabular}%
,\,\mathbb{T}_{9}^{(2)}=%
\begin{tabular}{l|l|l}
1 & 2 & 3 \\ \hline
4 & 5 & 7 \\ \hline
6 & 8 & 9%
\end{tabular}%
,\,\mathbb{T}_{9}^{(3)}=%
\begin{tabular}{l|l|l}
1 & 2 & 3 \\ \hline
4 & 5 & 8 \\ \hline
6 & 7 & 9%
\end{tabular}
\\
\,\mathbb{T}_{9}^{(4)} &=&%
\begin{tabular}{l|l|l}
1 & 2 & 3 \\ \hline
4 & 6 & 7 \\ \hline
5 & 8 & 9%
\end{tabular}%
,\,\mathbb{T}_{9}^{(5)}=%
\begin{tabular}{l|l|l}
1 & 2 & 3 \\ \hline
4 & 6 & 8 \\ \hline
5 & 7 & 9%
\end{tabular}%
,\,\mathbb{T}_{9}^{(6)}=%
\begin{tabular}{l|l|l}
1 & 2 & 4 \\ \hline
3 & 5 & 6 \\ \hline
7 & 8 & 9%
\end{tabular}
\\
\,\mathbb{T}_{9}^{(7)} &=&%
\begin{tabular}{l|l|l}
1 & 2 & 5 \\ \hline
3 & 4 & 6 \\ \hline
7 & 8 & 9%
\end{tabular}%
,\,\mathbb{T}_{9}^{(8)}=%
\begin{tabular}{l|l|l}
1 & 2 & 4 \\ \hline
3 & 5 & 7 \\ \hline
6 & 8 & 9%
\end{tabular}%
,\,\mathbb{T}_{9}^{(9)}=%
\begin{tabular}{l|l|l}
1 & 2 & 4 \\ \hline
3 & 5 & 8 \\ \hline
6 & 7 & 9%
\end{tabular}
\\
\,\mathbb{T}_{9}^{(10)} &=&%
\begin{tabular}{l|l|l}
1 & 2 & 5 \\ \hline
3 & 4 & 7 \\ \hline
6 & 8 & 9%
\end{tabular}%
,\,\mathbb{T}_{9}^{(11)}=%
\begin{tabular}{l|l|l}
1 & 2 & 5 \\ \hline
3 & 4 & 8 \\ \hline
6 & 7 & 9%
\end{tabular}%
,\,\mathbb{T}_{9}^{(12)}=%
\begin{tabular}{l|l|l}
1 & 2 & 6 \\ \hline
3 & 4 & 8 \\ \hline
5 & 7 & 9%
\end{tabular}
\\
\,\mathbb{T}_{9}^{(13)} &=&%
\begin{tabular}{l|l|l}
1 & 2 & 6 \\ \hline
3 & 4 & 7 \\ \hline
5 & 8 & 9%
\end{tabular}%
,\,\mathbb{T}_{9}^{(14)}=%
\begin{tabular}{l|l|l}
1 & 2 & 7 \\ \hline
3 & 4 & 8 \\ \hline
5 & 6 & 9%
\end{tabular}%
,\,\mathbb{T}_{9}^{(15)}=%
\begin{tabular}{l|l|l}
1 & 2 & 4 \\ \hline
3 & 6 & 7 \\ \hline
5 & 8 & 9%
\end{tabular}
\\
\mathbb{T}_{9}^{(16)} &=&%
\begin{tabular}{l|l|l}
1 & 2 & 4 \\ \hline
3 & 6 & 8 \\ \hline
5 & 7 & 9%
\end{tabular}%
,\,\mathbb{T}_{9}^{(17)}=%
\begin{tabular}{l|l|l}
1 & 2 & 6 \\ \hline
3 & 5 & 8 \\ \hline
4 & 7 & 9%
\end{tabular}%
,\,\mathbb{T}_{9}^{(18)}=%
\begin{tabular}{l|l|l}
1 & 2 & 6 \\ \hline
3 & 5 & 7 \\ \hline
4 & 8 & 9%
\end{tabular}
\\
\,\mathbb{T}_{9}^{(19)} &=&%
\begin{tabular}{l|l|l}
1 & 2 & 7 \\ \hline
3 & 5 & 8 \\ \hline
4 & 6 & 9%
\end{tabular}%
,\,\mathbb{T}_{9}^{(20)}=%
\begin{tabular}{l|l|l}
1 & 2 & 5 \\ \hline
3 & 6 & 8 \\ \hline
4 & 7 & 9%
\end{tabular}%
,\,\mathbb{T}_{9}^{(21)}=%
\begin{tabular}{l|l|l}
1 & 2 & 5 \\ \hline
3 & 6 & 7 \\ \hline
4 & 8 & 9%
\end{tabular}%
\end{eqnarray*}

For this, and larger dimensions, the minimum (and maximum) classical mutual information values are not known, there is not a unique table that defines these because it depends on the spectrum $\Lambda$.

\subsection{\label{secSeeSaw}A Partial Ordering on the Young Tableaux and the ``See-Saw'' Effect}

We cannot say whether the sum $H\left( \mathbf{a}%
\right) +H\left( \mathbf{b}\right) $ will increase from one table to
another, nevertheless it is possible in some cases to determine how the individual entropies will
change.
In the following, we assume \textit{a priori} majorization relations between vectors, that is partial ordering in the general case without reference to a specific probability distribution $\Lambda$.

\textit{Lemma}
If two $d_{A}\times d_{B} = d$ Young tableaux $\mathbb{T}^{(\mathcal{I})}_d,$ $\mathbb{T%
}^{(\mathcal{J})}_d\in Y_{d}$ are related by one transposition%
\begin{equation*}
\mathbb{T}^{(\mathcal{J})}_d=\pi\left( rs:tu\right) \mathbb{T}^{(\mathcal{I})}_d
\end{equation*}%
where $\pi\left( rs:tu\right) $ swaps the elements $\tau _{rs}$ and $%
\tau _{tu}$ of table $\mathbb{T}^{(\mathcal{I})}_d,$ then either $$\mathbf{a}^{(\mathcal{I})}\prec
\mathbf{a}^{(\mathcal{J})}, \, \mathbf{b}^{(\mathcal{I})}\succ \mathbf{b}^{(\mathcal{J})}$$ if $\delta >0$ or
$$\mathbf{a}^{(\mathcal{I})}\succ \mathbf{a}^{(\mathcal{J})}, \,\mathbf{b}^{(\mathcal{I})}\prec \mathbf{b}^{(\mathcal{J})}$$
if $\delta <0$ , where $\mathbf{a}^{(\mathcal{X})}$ and $\mathbf{b}^{(\mathcal{X})}$ are the row and
column MPVs of table $\mathbb{T}^{(\mathcal{X})},$ $\mathcal{X}\in \left\{ \mathcal{I},\mathcal{J}\right\} $, and $%
\delta =\tau _{tu}-\tau _{rs}.$ Say $r \geq t$ in the following, without loss of generality.

\textit{Proof}
The swap operator $\pi\left( rs:tu\right)$ cannot arbitrarily swap any two nonequal elements of a Young
tableau and stay within the set $Y_{d}.$ For a valid swap, it is necessary
that
\begin{eqnarray}
\label{swapcond1}
r &>&t  \label{ValidSeeSawSwapCond} \\
\label{swapcond2}
s &<&u  \notag
\end{eqnarray}%
If either $r=t$ or $s=u,$ then the swapping occurs between elements in the
same row or column, which is prohibited because it would violate the
Young tableau element ordering, similarly if $r>t$ and $s > t$. Note, however, that not all swaps
which satisfy equation (\ref{ValidSeeSawSwapCond}) are valid swaps, for example when $d_A = d_B = 3$, swapping $\tau_{12}$ and $\tau_{21}$ in table $\mathbb{T}^{1}$ concurs with these conditions but generates a table that is not in Young tableaux form. Therefore conditions (\ref{swapcond1}) are necessary but not sufficient.

Consider tables $\mathbb{T}^{ (\mathcal{I})}_d,$ $\mathbb{T}_d^{( \mathcal{J})}=\pi\left( rs:tu\right)
\mathbb{T}^{ (\mathcal{I})}_d$ $\in Y_{d}$. Two components of the row and column MPVs of $\mathbb{T}^{ (\mathcal{I})}_d$ are affected by the swap, they are%
\begin{eqnarray*}
 a^{(\mathcal{I})}_{r} &=&\tau _{r1}+\dots +\tau _{rs}+\dots +\tau
_{rd_{B}} \\
a^{(\mathcal{I})}_{t} &=&\tau _{t1}+\dots +\tau _{tu}+\dots +\tau
_{td_{B}} \\
b^{(\mathcal{I})}_{s} &=&\tau _{sj}+\dots +\tau _{rs}+\dots +\tau
_{d_{A}s} \\
b^{(\mathcal{I})}_{u} &=&\tau _{ul}+\dots +\tau _{tu}+\dots +\tau
_{d_{A}u}
\end{eqnarray*}%
After the swap, the MPVs of $\mathbb{T}^{(\mathcal{J})}_d,$  can be written
in terms of $\mathbf{a}^{(\mathcal{I})}$ and $\mathbf{b}^{(\mathcal{I})}$ components
\begin{eqnarray*}
a^{(\mathcal{J})}_{r} &=&\tau _{r1}+\dots +\tau _{tu}+\dots +\tau
_{rd_{B}}=a^{(\mathcal{I})}_{r}+\delta \\
a^{(\mathcal{J})}_{t} &=&\tau _{t1}+\dots +\tau _{rs}+\dots +\tau
_{td_{B}}=a^{(\mathcal{I})}_{t}-\delta \\
b^{(\mathcal{J})}_{s} &=&\tau _{1s}+\dots +\tau _{tu}+\dots +\tau
_{d_{A}s}=b^{(\mathcal{J})}_{s}+\delta \\
b^{(\mathcal{J})}_{u} &=&\tau _{1u}+\dots +\tau _{rs}+\dots +\tau
_{d_{A}u}=b^{(\mathcal{I})}_{u}-\delta
\end{eqnarray*}%
where $\delta =\tau _{tu}-\tau _{rs}$ and \ $\left\vert \delta \right\vert
<a^{(\mathcal{I})}_{i},b^{(\mathcal{I})}_{j}$ and $i=1,\dots ,d_{A},$ $%
j=1,\dots ,d_{B}$. The latter condition on the magnitude of $\delta $ ensures that
adding and subtracting it to the components of the MPVs
preserves their order; this is important when comparing
the majorization relation between two vectors. If the elements of $\mathbf{a}%
^{(\mathcal{I})}$ and $\mathbf{b}^{(\mathcal{I})}$ are ordered in decreasing order, then the
elements of $\mathbf{a}^{(\mathcal{J})}$ and $\mathbf{b}^{(\mathcal{J})}$ will be also.
This is a consequence of the fact that we only consider swaps which are operations $f: Y_d \rightarrow Y_d$ and more importantly that the arrangement of elements in the Young tables ensure the MPV elements are ordered; for row MPVs $a_1 \geq a_2 \geq \dots a_{d_A}$ and similarly for column MPVs.

Collating the aformentioned facts, it can be shown that the MPVs satisfy the majorization relations
\begin{equation*}
\mathbf{a}^{(\mathcal{J})}\prec \mathbf{a}^{(\mathcal{I})},\mathbf{b}^{(\mathcal{J})}\succ \mathbf{b}^{(\mathcal{I})}
\end{equation*}%
if $\delta >0$ and%
\begin{equation*}
\mathbf{a}^{(\mathcal{I})}\prec \mathbf{a}^{(\mathcal{J})},\mathbf{b}^{(\mathcal{I})}\succ \mathbf{b}^{(\mathcal{J})}
\end{equation*}%
if $\delta <0$. $\square$

\bigskip
Since entropy is Schur concave, this means that for, say, $\delta >0$
\begin{equation*}
H\left( \mathbf{a}^{(\mathcal{J})}\right) \geq H\left( \mathbf{a}^{(\mathcal{I})}\right) ,\\
H\left( \mathbf{b}^{(\mathcal{J})}\right) \leq H\left( \mathbf{b}^{(\mathcal{I})}\right)
\end{equation*}%
and because a transposition causes one entropy to increase and the other to decrease in this way, we call this the ``see-saw'' effect.

\begin{figure}[h!]
\includegraphics[width=3.5in]{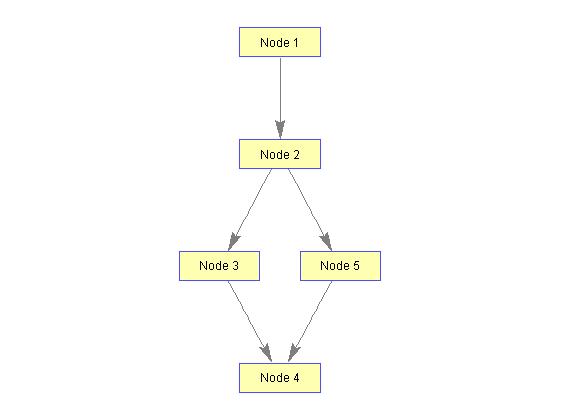}
\caption{Connected and directed graph $G_{row}\left( Y_{6}\right)$, each node represent a table, e.g. Node 1 is table $\mathbb{T}_6^{(1)}$ in section \ref{sec2x3Tables}.}
\label{fig2x3YTGraph}
\end{figure}

\begin{figure}[h!]
\includegraphics[width=3.5in]{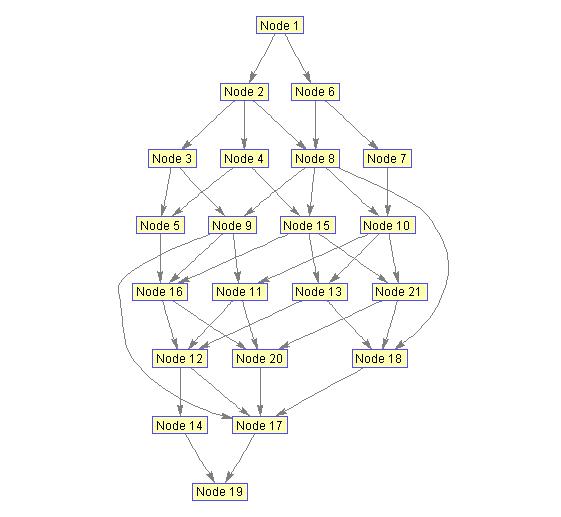}
\caption{Connected and directed graph $G_{row}\left( Y_{9}\right)$, each node represent a table, e.g. Node 1 is table $\mathbb{T}_9^{(1)}$ in section \ref{sec3x3Tables}. It would be symmetrical like $G_{row}\left( Y_{6}\right)$ if the transposes of the tables in $Y_9$ were included.}
\label{fig3x3YTGraph}
\end{figure}

It is possible to represent this effect by a directed graph $G=G\left( Y_{d}\right) $.
The nodes correspond to the tables and the edges connect two tables that are
related by a swap. A graph $G$ can only represent row or column MPV majorization; we call the two types of
graphs $G_{row}$ and $G_{col}.$ The direction of the edge
connecting two nodes represents an MPV majorization relation, hence an arrow
``$\rightarrow $'' from node $\mathcal{I}$ to $\mathcal{J}$ on the graph $G_{row}$ implies $%
\mathbf{a}^{(\mathcal{I})}\succ \mathbf{a}^{(\mathcal{J})}$. From the lemma above this also
means that $\mathbf{b}^{(\mathcal{I})}\prec \mathbf{b}^{(\mathcal{J})}$ is also true. Hence on $G_{%
row}$, a ``$\rightarrow $'' arrow will exist between nodes $\mathcal{I}$ and $\mathcal{J},$
whereas on $G_{col}$ there will be a ``$\leftarrow $'' arrow between the
same two nodes. Hence $G_{row}$ and $G_{col}$ are equal up to
a change in their arrow directions, this allows us to just focus on either row
or column majorization since one graph is easily obtained from the other.
Figures \ref{fig2x3YTGraph} and \ref{fig3x3YTGraph} display the graphs $G_{row}\left( Y_{6}\right) $ and $G_{row}%
\left( Y_{9}\right) .$

Because of the transitivity property of majorization, if for tables
$\mathbb{T}^{(\mathcal{I})}$ , $\mathbb{T}^{(\mathcal{J})},$ $\mathbb{T}^{(\mathcal{K})}\in Y_{d}$ we have that $%
\mathbf{a}^{(\mathcal{I})}\succ \mathbf{a}^{(\mathcal{J})}$ and $\mathbf{a}^{(\mathcal{J})}\succ \mathbf{a}^{(\mathcal{K})},$
then $\mathbf{a}^{(\mathcal{I})}\succ \mathbf{a}^{(\mathcal{K})}.$ The graph $G_{row}\left(
Y_{d}\right) $ captures this more general majorization relation.
Following a sequence of nodes connected by edges all directed the same way
gives a subset of tables that obey this partial
order. For example, in the graph $G_{row}\left(
Y_{6}\right) ,$ a set of connected nodes is Node 1 $\rightarrow $ Node 2 $%
\rightarrow $ Node 3 $\rightarrow $ Node 4 from which we can say, for instance, that $\mathbf{a}%
^{1}\succ \mathbf{a}^{4}$ \textit{and} $\mathbf{b}%
^{1}\prec \mathbf{b}^{4}$. Except for Nodes 3 and 5, all other pairs of
nodes in $G_{row}\left( Y_{6}\right) $ are connected by arrows all
pointing in the same direction hence we can determine almost all the column and row
MPV majorization relations of any two tables in $Y_{6}.$ Upon inspection of the tables corresponding to Nodes 3 and 5, we see their MPVs are incomparable, which is reflected by the fact they are unconnected in $G(Y_6)$.
Hence we can make the following statement:

\textit{Two tables }$\mathbb{T}^{(\mathcal{I})}$\textit{\ and }$\mathbb{T}^{(\mathcal{J})}$\textit{\ in }$Y_{d},$%
\textit{\ represented by nodes }$N^{(\mathcal{I})}$\textit{\ and }$N^{(\mathcal{J})}$\textit{\ in
the graph }$G_{row}\left( Y_{d}\right) ,$\textit{\ have a
majorization relation between their row and column MPVs if }$N^{(\mathcal{I})}$\textit{%
\ and }$N^{(\mathcal{J})}$\textit{\ are connected by arrows all pointing in the same
direction.}

\subsection{\label{secSpecCase}Special Case Results}

In some cases we can use simple mathematics to calculate whether the mutual information will increase or decrease from one permutation to another. Let us denote a probability vector with $n$ elements $\mathbf{p}=(p_1,p_2,\dots,p_n)$ where $p_1\geq p_2 \geq \dots \geq p_n$. Two results follow.

\bigskip

\textit{Lemma 1} Given
\begin{eqnarray*}
\mathbb{T}_{n}^{(1)}=%
\begin{tabular}{l|l|l}
$p_1$ & $\dots$ & $p_k$ \\ \hline
$p_{k+1}$ & $\dots$ & $\dots$ \\ \hline
$\dots$ & $\dots$ & $p_n$
\end{tabular}%
\\
\mathbb{T}_{n}^{(2)}=%
\begin{tabular}{l|l|l}
$p_1$ & $\dots$ & $p_{k+1}$ \\ \hline
$p_{k}$ & $\dots$ & $\dots$ \\ \hline
$\dots$ & $\dots$ & $p_n$%
\end{tabular}%
\end{eqnarray*}
with $k < n$, and $p_i=0$ for all $i>k$.
Then $I(\mathbb{T}_{n}^{(1)}) \leq I(\mathbb{T}_{n}^{(2)})$.

\bigskip

\textit{Proof 1} Let $h(x) = -x \log x$. Then (ignoring the total entropy contribution to the mutual information)
\begin{align}
I(\mathbb{T}_{n}^{(1)}) = \sum_{i=1}^k h(p_i)
\end{align}
\begin{align}
I(\mathbb{T}_{n}^{(2)}) = h(p_1+p_k)+ \sum_{i=2}^{k-1} h(p_i) + h\left(\sum_{i=1}^{k-1}p_i\right) + h(p_k)
\end{align}
hence it must be shown that
\begin{align}
I({T}_{n}^{(2)})-I({T}_{n}^{(1)})\geq 0
\end{align}
or
\begin{align}
h(p_1+p_k) + h(1 - p_k) - h(p_1)\geq 0
\end{align}
Writing this as $h(p_1+p_k) - h(p_1) + h(1 - p_k) - h(1)$ and using the mean value theorem $f(b) - f(a) = (b-a)f'(c)$ where $c \in \left[a,b\right]$ then the condition becomes
\begin{align}
h'(p_1 + \beta) - h'(1-\alpha)\geq 0
\end{align}
where $\alpha, \beta \in \left[0,p_k\right]$. The derivative of the function is $h'(x) = -k_1 \ln x - k_2$ where $k_1,k_2$ are constants so we obtain
\begin{align}
\ln\frac{1-\alpha}{p_1 + \beta}\geq 0
\end{align}
or
\begin{align}
1-\alpha \geq p_1 + \beta
\end{align}
In the worst case (minimum RHS and maximum LHS) the inequality becomes
\begin{align}
1-p_k \geq p_1
\end{align}
which is true since $1 = \sum_{i=1}^k p_i \geq p_1 + p_k$. $\square$

\bigskip

\textit{Lemma 2}
If we have
\begin{eqnarray*}
\mathbb{T}_{n}^{(1)}=%
\begin{tabular}{l|l|l}
$p_1$ & $\dots$ & $p_k$ \\ \hline
$p_{k+1}$ & $\dots$ & $\dots$ \\ \hline
$\dots$ & $\dots$ & $p_n$
\end{tabular}%
\\
\mathbb{T}_{n}^{(2)}=%
\begin{tabular}{l|l|l}
$p_1$ & $\dots$ & $p_{k+1}$ \\ \hline
$p_{k}$ & $\dots$ & $\dots$ \\ \hline
$\dots$ & $\dots$ & $p_n$%
\end{tabular}%
\end{eqnarray*}
with $k < n$, and $p_i=0$ for all $i>k+1$.
Then $I(\mathbb{T}_{n}^{(1)}) \leq I(\mathbb{T}_{n}^{(2)})$.

\bigskip

\textit{Proof 2}
The condition
\begin{align}
I(\mathbb{T}_{n}^{(2)})-I(\mathbb{T}_{n}^{(1)})\geq 0
\end{align}
is equivalent to
\begin{align}
h(p_1+p_k) - h(p_1+p_{k+1}) + h(1-p_k) - h(1-p_{k+1}) \geq 0
\end{align}
Using the mean value theorem again we obtain
\begin{align}
h'(p_1 + p_{k+1} + \gamma) - h'(1-p_{k} + \delta)\geq 0
\end{align}
where $\gamma, \delta \in \left[0,p_k-p_{k+1}\right]$, so
\begin{align}
\ln\frac{1-p_{k} + \delta}{p_1 + p_{k+1} + \gamma}\geq 0
\end{align}
or
\begin{align}
1-p_{k} + \delta \geq p_1 + p_{k+1} + \gamma
\end{align}
In the worst case (minimum RHS and maximum LHS) the inequality becomes
\begin{align}
 1-p_{k}\geq p_1 + p_{k}
\end{align}
which is true since $1 = \sum_{i=1}^{k+1} p_i \geq p_1 + p_{k-1} + p_k \geq p_1 + 2p_k$. $\square$

\section{Conclusion}

Intuitively it would seem right that the state on a unitary orbit that is the maximally (minimally) correlated is the one which is a convex combination of maximum (minimum) QMI pure states. Here we have formally shown that this is true, however it is highly non-trivial and additional considerations beyond majorization are required. The entire analysis boils down to the dependence of the mutual information on orderings of elements in Young tableaux, and these tables display a subtle property in terms of a partial order with respect to marginal entropies.

This work finds application to closed system thermodynamics when one includes the additional constraint of energy conservation.

Many challenging questions remain, the obvious one being, given a spectrum of a bipartite system, what is the arrangement of its eigenvalues that gives the minimum (and maximum classical) mutual information? Can anything be gleaned from our analysis about other correlation/entanglement/concave functions? What about the inverse of the problem, varying the QMI when the marginals are fixed but the global state is not (but must be compatible with the local description)?

In terms of thermodynamical work, can we devise a simple collision model in which correlations can remain non-zero but equilibration still occurs? Can our work be connected to recent advancements in the resource theory of athermal states \cite{TDResource}? What do our results contribute to traditional open systems quantum thermodynamics?

We see that there are many avenues that our investigations could lead on to, and hope that our work provides use to people investigating the thermodynamics of quantum systems.

\begin{acknowledgments}
We would like to thank the two anonymous referees of Ref. \cite{FirstPaper} for stimulating a discussion on the physical assumptions of thermodynamics in a closed system. This work was supported by EPSRC. 
\end{acknowledgments}

\appendix

\section{Details on the construction of marginal spectras}

For two qubits, the minimum QMI state, $\rho_{min}$, must correspond to a point on the line $g_p, \, p=1,2$ in figure \ref{figLALBfgh}. There are two cases to consider.

\subsection{$p=1$, $\lambda_2 = \lambda_3$}

The inequality in equation \eqref{ineq3} reduces to $$ \lambda_A + \lambda_B \geq 2(\lambda_3 + \lambda_4) $$ and it is now redundant because it is no longer independent - it can be formed by adding inequalities \eqref{ineq1} and \eqref{ineq2}. Therefore it places no additional constraint on $\mathcal{R}$ and the boundary closest to the origin is in fact a corner where the lines $f_1$ and $f_2$ intersect. The minimum values of $\lambda_A$ and $\lambda_B$ occur at this intersection, where $\lambda_A = \lambda_B = \lambda_3 + \lambda_4$.

\subsubsection{\label{g2}$p=2$, $\lambda_2 > \lambda_3$}

Now $$ \lambda_A + \lambda_B >  2(\lambda_3 + \lambda_4) $$ and this inequality is a true constraint on $\mathcal{R}$. The line $g_2$ cuts the bottom left corner of $\mathcal{R}$ and the minimum of $H(\lambda_A) + H(\lambda_B)$ is somewhere along it. Let us define $$K =  \lambda_2 + \lambda_3 + 2\lambda_4$$ then $g_2$ can be rewritten $\lambda_B = K - \lambda_A$ and the function to be minimised on this line is $H(\lambda_A) + H(K - \lambda_A)$, with $\lambda_A \in [0,K]$. The domain is now convex and the function concave
and symmetric about the maximum at $\lambda_A = \frac{K}{2}$ (where $\lambda_A=\lambda_B$, the symmetry line of $\mathcal{R}$) hence the minima occur at the edges of $g_2$.

Let us call the coordinates of the boundaries of $g_2$ $R_1,R_2 \in \mathcal{R}$, shown in figure \ref{figLALBfgh}. 
By symmetry we need only consider one of these points, $R_1$ say ($R_2$ is given by swapping $\lambda_A, \lambda_B$ in $R_1$), its coordinate is either at the intersection of the lines $h_1$ and $g_2$ or $f_2$ and $g_2$. $f_2$ and $g_2$ meet at $(\lambda_A,\lambda_B) = (\lambda_2+\lambda_4,\lambda_3+\lambda_4)=:(\lambda^*_A,\lambda^*_B)$. However there are two situations to consider for the intersection of $g_2$ and $h_1$ depending on min$\{\lambda_1 - \lambda_3,\lambda_2-\lambda_4 \}$.\newline\newline
If $h_1=\textrm{min}\{\lambda_1 - \lambda_3,\lambda_2-\lambda_4 \} = \lambda_1 - \lambda_3$: \newline
then $g_2$ and $h_1$ meet at $(\lambda_A,\lambda_B) $
\begin{align*}
&=(\frac{1}{2}(\lambda_1-\lambda_2)+\lambda_2+\lambda_4,\lambda_3+\lambda_4-\frac{1}{2}(\lambda_1-\lambda_2))\\
&=:(\lambda'_A,\lambda'_B).
\end{align*}
\newline\newline
If $h_1 = \min\{\lambda_1 - \lambda_3,\lambda_2-\lambda_4 \} = \lambda_2 - \lambda_4$:\newline
then  $g_2$ and $h_1$ meet at $(\lambda_A,\lambda_B) $
\begin{align}
&=(\frac{1}{2}(\lambda_3-\lambda_4)+\lambda_2+\lambda_4,\lambda_3+\lambda_4-\frac{1}{2}(\lambda_3-\lambda_4))\\ &=:(\lambda''_A,\lambda''_B).
\end{align}
\newline\newline
It is clear that $\lambda'_A,\lambda''_A \geq \lambda^*_A$ and $\lambda'_B,\lambda''_B \leq \lambda^*_B$. This implies that in both cases $h_1$ intersects $g_2$ at a point that is further from the diagonal $\lambda_A = \lambda_B$ than the intersection of $f_2$ and $g_2$ and so the coordinates $(\lambda'_A,\lambda'_B),(\lambda''_A,\lambda''_B)$ are not in $\mathcal{R}$, hence $R_1 = (\lambda^*_A,\lambda^*_B)$.

Therefore, the minimum of $H(\lambda_A) + H(\lambda_B)$ occurs at $R_1$ where
\begin{align}
\label{lambdaA_min1}
\lambda_A = \lambda_2 + \lambda_4\\
\label{lambdaB_min1}
\lambda_B = \lambda_3 + \lambda_4
\end{align}
and at $R_2$
\begin{align}
\label{lambdaA_min2}
\lambda_A = \lambda_3 + \lambda_4\\
\label{lambdaB_min2}
\lambda_B = \lambda_2 + \lambda_4
\end{align}
It is evident now that $g_1$ in the previous is a special case of this when $\lambda_2 = \lambda_3$.

\bigskip

The only task left is to determine the composite state $\rho_{min}$ which has reduced states with the eigenvalues given in equations \eqref{lambdaA_min1} and \eqref{lambdaB_min1} (or \eqref{lambdaA_min2} and \eqref{lambdaB_min2}).
The inequalities \eqref{ineq1} - \eqref{ineq3} which give the lines $f_p,g_p$ are obtained by using the relation
\begin{eqnarray}
\label{inf}
\inf_{\tau \in \mathcal{D}(\mathcal{H},4,\lambda)} \tr(O\tau)= \lambda_1 O_4 +\lambda_2 O_3+\lambda_3 O_2+\lambda_4 O_1
\end{eqnarray}
where $O$ is an arbitrary two-qubit hermitian operator with eigenvalues $O_1 \geq O_2 \geq O_3 \geq O_4$ and $\tau$ is four dimensional state with eigenvalues $\Lambda =\{\lambda_i \}^4_{i=1}$ ordered in the same way. This infimum is reached when $O$ and $\tau$ are diagonal in the same basis with their eigenvalues arranged in such a way that equation \eqref{inf} is obtained. 

Let us write the marginals of a two qubit state $\rho$ as
\begin{align}
\rho_A &= \lambda_A \Pi_A + (1-\lambda_A)(I-\Pi_A)\\
\rho_B &= \lambda_B \Pi_B + (1-\lambda_B)(I-\Pi_B)
\end{align}
and $\Pi_A, \Pi_B$ are the projectors onto the eigenstates of $\rho_A, \rho_B$ associated with the lowest eigenvalue.

As noted above, the point $R_1$ in $\mathcal{R}$ corresponding to the minimum QMI state is an intersection of two boundary lines: $\lambda_A = \lambda_3 + \lambda_4$ and $\lambda_A + \lambda_B = \lambda_2 + \lambda_3 + 2\lambda_4$. Hence $\rho_{min}$ must commute with the two operators $O, O'$ that give rise to these lines.

To obtain inequality \eqref{ineq2} we set  $O = \Pi_A$,
then $\{O_1,O_2,O_3,O_4 \} = \{1,1,0,0\}$ and
\begin{align}
\lambda_A &= \tr(\rho \Pi_A)\\
&\geq \inf_{\tau \in \mathcal{D}(\mathcal{H},4,\lambda)} \tr(\tau \Pi_A)\\
&=\lambda_4 + \lambda_3
\end{align}

Similarly, to obtain inequality \eqref{ineq3} we set $O' = \Pi_A+\Pi_B$. Let us suppose, without loss of generality, that $\Pi_A = \ket{0}\bra{0}$ and $\Pi_B = \ket{v}\bra{v}$, where $\ket{v} = U_B\ket{0} $ and $U_B$ is a single qubit unitary. $O$ and $U_B^{\dag}OU_B$ are cospectral and have eigenvalues $\{O_1,O_2,O_3,O_4 \} = \{2,1,1,0\}$. 
Thus
\begin{align}
\lambda_A + \lambda_B &= \tr(\rho (\Pi_A+\Pi_B)) \\
&\geq \inf_{\tau \in \mathcal{D}(\mathcal{H},4,\lambda)} \tr(\tau (\Pi_A+\Pi_B))\\
 &= 2\lambda_4 + \lambda_3 + \lambda_2
\end{align}
\newline

Without loss of generality $O, O'$ can be set to be diagonal in the same, say computational, basis.
In matrix form they are

\begin{eqnarray}
O=
\left(
  \begin{array}{cccc}
    1 & 0 & 0 & 0 \\
    0 & 1 & 0 & 0 \\
    0 & 0 & 0 & 0 \\
    0 & 0 & 0 & 0 \\
  \end{array}
\right)\\
O'=
\left(
  \begin{array}{cccc}
    2 & 0 & 0 & 0 \\
    0 & 1 & 0 & 0 \\
    0 & 0 & 1 & 0 \\
    0 & 0 & 0 & 0 \\
  \end{array}
\right)
\end{eqnarray}

Then $\rho_{min}$ must satisfy $\left[\rho_{min},O\right] = 0$ and $\left[\rho_{min},O'\right] = 0$. Alternatively, this is fulfilled if we require $\left[\rho_{min},O + O'\right] = 0$ and $O \neq -O'$ (which is true). Given
\begin{eqnarray}
O+O'=
\left(
  \begin{array}{cccc}
    3 & 0 & 0 & 0 \\
    0 & 2 & 0 & 0 \\
    0 & 0 & 1 & 0 \\
    0 & 0 & 0 & 0 \\
  \end{array}
\right),
\end{eqnarray}
then $\rho_{min}$ must be diagonal in the same, computational, basis. Since $\mathrm{tr}(\rho_{min} \Pi_A)= \lambda_3 + \lambda_4$ then
\begin{eqnarray}
\label{eqrhomin}
\rho_{min}=
\left(
  \begin{array}{cccc}
    \lambda_1 & 0 & 0 & 0 \\
    0 & \lambda_2 & 0 & 0 \\
    0 & 0 & \lambda_3 & 0 \\
    0 & 0 & 0 & \lambda_4 \\
  \end{array}
\right)
\end{eqnarray}

This state and its equivalence class is a classically correlated state, and it is the minimiser for the QMI.

\section{Maximum $\Delta I_E$}

To see more clearly that equation (\ref{EconsThL}) describes a line let us define $\lambda_A = y$, $\lambda_B = x$, $\cos \theta_A = t$, $\cos \theta_B = s$ and $C = 1-E \geq 0$. Then the equation can be recast in terms of these variables
$$
yt + xs = \frac{1}{2}(s+t) - C
$$
We call this line $F(s,t)$. A special case occurs for $F(0,0)$ where $E=1$ and no other energy is allowed. These lines have symmetry about the lines $x=\frac{1}{2}, y = \frac{1}{2}$.

The region of states with initial and final energies equal to $E$ will be where the lines $F(s,t)$ intersect the region $\mathcal{R}$.
There are four important lines that frame the set of all $F(s,t)$ and they occur when $s=\pm 1, t=\pm 1$. The intersection of these lines, denoted by $F(s,t) \, \& \, F(s',t')$, is given below in terms of the coordinate representation $(x,y)$
\begin{align}
\label{F1}
F(1,1) \, \& \, F(1,-1) &: \left(\frac{1}{2},\frac{1}{2}-C\right)\\
\label{F2}
F(1,1) \, \& \, F(-1,1) &: \left(\frac{1}{2}-C,\frac{1}{2}\right)\\
\label{F3}
F(-1,-1) \, \& \, F(1,-1) &: \left(\frac{1}{2}+C,\frac{1}{2}\right)\\
\label{F4}
F(-1,-1) \, \& \, F(-1,1) &: \left(\frac{1}{2},\frac{1}{2}+C\right)
\end{align}

\begin{centering}
\begin{figure}[h!]
\includegraphics[width=3.8in]{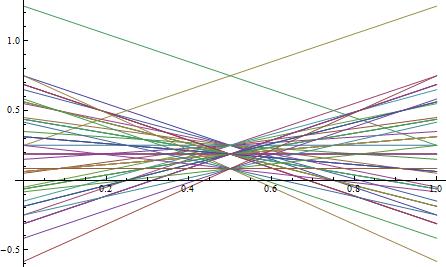}
\caption{The region occupied by the lines $F[s,t]$ is the part outside the central diamond; this is the region of states which have energy $E$ for some angles $\theta_A, \theta_B$. The diamond is bounded by the four lines in equations \eqref{F1} to \eqref{F4}, and $E=0.75$. Some typical lines in $F[s,t]$ are added, they never go inside the diamond and the system is symmetric about the lines  $x=\frac{1}{2}, y=\frac{1}{2}$.}
\label{Fst_region}
\end{figure}
\end{centering}

This is depicted in figure \ref{Fst_region}. A general line in $F(s,t)$ passes through the points $(\frac{1}{2},\frac{1}{2}-\frac{C}{t})$ and $(\frac{1}{2}-\frac{C}{s},\frac{1}{2})$ and so it never goes inside the region bounded by the lines  $F(1,1),F(1,-1),F(-1,1),F(-1,-1)$. This is also shown in the figure.

The region $\mathcal{R}$ is restricted to $x, y \leq \frac{1}{2}$. The set of lines $F(s,t)$ that fall into this range are $F(1,1)$ and all the lines ``below'' it, that is all the lines satisfying $x + y \leq 1 - C$.

\bibliography{twoquDitref}

\end{document}